\documentclass[11pt]{article}\usepackage[]{graphicx}\usepackage[]{color}
\makeatletter
\def\maxwidth{ %
  \ifdim\Gin@nat@width>\linewidth
    \linewidth
  \else
    \Gin@nat@width
  \fi
}
\makeatother

\definecolor{fgcolor}{rgb}{0.345, 0.345, 0.345}
\newcommand{\hlnum}[1]{\textcolor[rgb]{0.686,0.059,0.569}{#1}}%
\newcommand{\hlstr}[1]{\textcolor[rgb]{0.192,0.494,0.8}{#1}}%
\newcommand{\hlcom}[1]{\textcolor[rgb]{0.678,0.584,0.686}{\textit{#1}}}%
\newcommand{\hlopt}[1]{\textcolor[rgb]{0,0,0}{#1}}%
\newcommand{\hlstd}[1]{\textcolor[rgb]{0.345,0.345,0.345}{#1}}%
\newcommand{\hlkwa}[1]{\textcolor[rgb]{0.161,0.373,0.58}{\textbf{#1}}}%
\newcommand{\hlkwb}[1]{\textcolor[rgb]{0.69,0.353,0.396}{#1}}%
\newcommand{\hlkwc}[1]{\textcolor[rgb]{0.333,0.667,0.333}{#1}}%
\newcommand{\hlkwd}[1]{\textcolor[rgb]{0.737,0.353,0.396}{\textbf{#1}}}%

\usepackage{framed}
\makeatletter
\newenvironment{kframe}{%
 \def\at@end@of@kframe{}%
 \ifinner\ifhmode%
  \def\at@end@of@kframe{\end{minipage}}%
  \begin{minipage}{\columnwidth}%
 \fi\fi%
 \def\FrameCommand##1{\hskip\@totalleftmargin \hskip-\fboxsep
 \colorbox{shadecolor}{##1}\hskip-\fboxsep
     \hskip-\linewidth \hskip-\@totalleftmargin \hskip\columnwidth}%
 \MakeFramed {\advance\hsize-\width
   \@totalleftmargin\z@ \linewidth\hsize
   \@setminipage}}%
 {\par\unskip\endMakeFramed%
 \at@end@of@kframe}
\makeatother

\definecolor{shadecolor}{rgb}{.97, .97, .97}
\definecolor{messagecolor}{rgb}{0, 0, 0}
\definecolor{warningcolor}{rgb}{1, 0, 1}
\definecolor{errorcolor}{rgb}{1, 0, 0}
\newenvironment{knitrout}{}{} 

\usepackage{alltt}
\usepackage{natbib,amssymb,amsmath,fullpage,hyperref}

\newcommand{\bl}[1]{{\mathbf #1}}
\newcommand{\bs}[1]{{\boldsymbol #1}}

\newcommand{\Var}[1]{{\text{Var}}[ \ensuremath{ #1 } ]  }
\newcommand{\Cov}[1]{{\text{Cov}}[ \ensuremath{ #1 } ]  }

\IfFileExists{upquote.sty}{\usepackage{upquote}}{}
\begin{document}

\title{Dyadic data analysis with  {\tt amen}}
\author{Peter D. Hoff
\thanks{Departments of Statistics and Biostatistics, University of Washington, Seattle. 
\url{http://www.stat.washington.edu/\~pdhoff/}
Development of this software and tutorial was supported  by
NIH grant R01HD067509. } }
\maketitle

\begin{abstract} 
Dyadic data on pairs of objects, such as relational  or social network data, 
%
often exhibit
strong statistical dependencies. Certain types of 
second-order dependencies, such as 
degree heterogeneity and reciprocity, can be well-represented with 
additive random effects models. 
Higher-order dependencies, 
such as transitivity and stochastic equivalence, can often be represented with  multiplicative effects. 
The {\tt amen} package for the {\sf R} statistical 
computing environment
provides estimation and inference 
for a class of additive and multiplicative random effects models 
for 
ordinal, continuous, binary and other types of 
dyadic data. 
The package also provides methods 
for missing, censored and fixed-rank nomination data, as well 
as longitudinal dyadic data.  This tutorial illustrates the 
{\tt amen} package 
via  example statistical analyses of several of these different  data types.

\smallskip


\end{abstract}

\tableofcontents

\section{The Gaussian AME model}
A pair of objects, individuals or nodes is called a \emph{dyad}, and 
a variable that is measured or observed on multiple dyads
is called a \emph{dyadic variable}.
Data on such a variable may be referred to as 
dyadic data, relational data, or 
network data (particularly if the variable is 
binary). 
Dyadic data for 
a population of  $n$  objects, individuals or nodes 
may be represented as 
a \emph{sociomatrix}, an $n\times n$  square matrix $\bl Y$ 
with an undefined diagonal. 
The $i,j$th entry of $\bl Y$, 
denoted  $y_{i,j}$, 
gives the 
value of the variable for  dyad $\{i,j\}$
from the perspective of node $i$, or in the direction 
from $i$ to $j$. 
For example, in a dataset describing friendship relations, 
$y_{i,j}$ might represent a quantification of 
how much person $i$ likes person $j$.  
A running example in this 
section will be  an analysis of international 
trade data,  where 
$y_{i,j}$ is the (log) dollar-value of exports from country $i$ to 
country $j$. These 
data can be obtained 
from the 
{\tt IR90s} dataset included in the {\tt amen} package.  
Specifically, we will analyze 
trade data between the
30 countries having the highest
GDPs:
\begin{knitrout}\footnotesize
\definecolor{shadecolor}{rgb}{0.969, 0.969, 0.969}\color{fgcolor}\begin{kframe}
\begin{alltt}
\hlcom{#### ---- obtain trade data from top 30 countries in terms of GDP}
\hlkwd{data}\hlstd{(IR90s)}

\hlstd{gdp}\hlkwb{<-}\hlstd{IR90s}\hlopt{$}\hlstd{nodevars[,}\hlnum{2}\hlstd{]}
\hlstd{topgdp}\hlkwb{<-}\hlkwd{which}\hlstd{(gdp}\hlopt{>=}\hlkwd{sort}\hlstd{(gdp,}\hlkwc{decreasing}\hlstd{=}\hlnum{TRUE}\hlstd{)[}\hlnum{30}\hlstd{] )}
\hlstd{Y}\hlkwb{<-}\hlkwd{log}\hlstd{( IR90s}\hlopt{$}\hlstd{dyadvars[topgdp,topgdp,}\hlnum{2}\hlstd{]} \hlopt{+} \hlnum{1} \hlstd{)}

\hlstd{Y[}\hlnum{1}\hlopt{:}\hlnum{5}\hlstd{,}\hlnum{1}\hlopt{:}\hlnum{5}\hlstd{]}
\end{alltt}
\begin{verbatim}
          ARG        AUL       BEL        BNG        BRA
ARG        NA 0.05826891 0.2468601 0.03922071 1.76473080
AUL 0.0861777         NA 0.3784364 0.10436002 0.21511138
BEL 0.2700271 0.35065687        NA 0.01980263 0.39877612
BNG 0.0000000 0.01980263 0.1222176         NA 0.01980263
BRA 1.6937791 0.23901690 0.6205765 0.03922071         NA
\end{verbatim}
\end{kframe}
\end{knitrout}

\subsection{The social relations model}
Dyadic
data often exhibit 
certain types of statistical dependencies.    
For example, it is often  the case that
observations in a given row of the sociomatrix are
similar to or correlated with each other. 
This should not
be too surprising, as these  observations all share a
common ``sender,'' or row index. 
If a sender $i_1$ is more ``sociable'' than sender $i_2$, we would 
expect the values in row $i_1$ to be larger than 
those in row $i_2$, on average. 
In this way, heterogeneity of the nodes in terms of their 
``sociability'' corresponds to a
large variance of the row means of the sociomatrix. 
Similarly, 
nodal heterogeneity in ``popularity'' corresponds to 
a large variance in the column means. 

A classical approach to evaluating across-row and across-column 
heterogeneity in a data matrix is the ANOVA decomposition. 
A model-based version of the ANOVA decomposition posits that 
the variability of 
the $y_{i,j}$'s around some overall mean is well-represented by 
additive row and column effects:
\[ 
y_{i,j} = \mu + a_i +b_j + \epsilon_{i,j}. 
\] 
In this model, 
heterogeneity among the parameters 
$\{a_i: i=1,\ldots, n\}$ and 
$\{b_j : j=1,\ldots, n\}$ corresponds to 
observed heterogeneity in the row means 
and column means of the sociomatrix, respectively. 
If the $\epsilon_{i,j}$'s are assumed to be 
i.i.d.\ from a mean-zero normal distribution, 
the hypothesis of no row heterogeneity 
(all $a_i$'s equal to zero) or 
no column heterogeneity (all $b_j$'s equal to zero) 
can be evaluated with normal-theory $F$-tests. 
For the trade data, this can be done in {\sf R} as follows:

\begin{knitrout}\footnotesize
\definecolor{shadecolor}{rgb}{0.969, 0.969, 0.969}\color{fgcolor}\begin{kframe}
\begin{alltt}
\hlcom{#### ---- ANOVA for trade data}

\hlstd{Rowcountry}\hlkwb{<-}\hlkwd{matrix}\hlstd{(}\hlkwd{rownames}\hlstd{(Y),}\hlkwd{nrow}\hlstd{(Y),}\hlkwd{ncol}\hlstd{(Y))}
\hlstd{Colcountry}\hlkwb{<-}\hlkwd{t}\hlstd{(Rowcountry)}

\hlkwd{anova}\hlstd{(}\hlkwd{lm}\hlstd{(} \hlkwd{c}\hlstd{(Y)} \hlopt{~} \hlkwd{c}\hlstd{(Rowcountry)} \hlopt{+} \hlkwd{c}\hlstd{(Colcountry) ) )}
\end{alltt}
\begin{verbatim}
Analysis of Variance Table

Response: c(Y)
               Df Sum Sq Mean Sq F value    Pr(>F)    
c(Rowcountry)  29 202.48  6.9819  29.524 < 2.2e-16 ***
c(Colcountry)  29 206.32  7.1144  30.084 < 2.2e-16 ***
Residuals     811 191.79  0.2365                      
---
Signif. codes:  0 '***' 0.001 '**' 0.01 '*' 0.05 '.' 0.1 ' ' 1
\end{verbatim}
\end{kframe}
\end{knitrout}

The results indicate a large degree of heterogeneity 
of the 
countries as both exporters and importers - much more than 
would be expected if the ``true'' $a_i$'s were all zero, 
or the ``true'' $b_j$'s were all zero  (and the 
$\epsilon_{i,j}$'s were i.i.d.). 
Based on this result, the next steps in a data analysis might 
include comparisons of the row means or of the column means, that is, 
comparisons of the countries in terms of their total or average 
imports and exports. This can equivalently be done via 
comparisons among estimates of the row and column effects:
\begin{knitrout}\footnotesize
\definecolor{shadecolor}{rgb}{0.969, 0.969, 0.969}\color{fgcolor}\begin{kframe}
\begin{alltt}
\hlcom{#### ---- comparison of countries in terms of row and column means}
\hlstd{rmean}\hlkwb{<-}\hlkwd{rowMeans}\hlstd{(Y,}\hlkwc{na.rm}\hlstd{=}\hlnum{TRUE}\hlstd{) ; cmean}\hlkwb{<-}\hlkwd{colMeans}\hlstd{(Y,}\hlkwc{na.rm}\hlstd{=}\hlnum{TRUE}\hlstd{)}

\hlstd{muhat}\hlkwb{<-}\hlkwd{mean}\hlstd{(Y,}\hlkwc{na.rm}\hlstd{=}\hlnum{TRUE}\hlstd{)}
\hlstd{ahat}\hlkwb{<-}\hlstd{rmean}\hlopt{-}\hlstd{muhat}
\hlstd{bhat}\hlkwb{<-}\hlstd{cmean}\hlopt{-}\hlstd{muhat}

\hlcom{# additive "exporter" effects}
\hlkwd{head}\hlstd{(} \hlkwd{sort}\hlstd{(ahat,}\hlkwc{decreasing}\hlstd{=}\hlnum{TRUE}\hlstd{)  )}
\end{alltt}
\begin{verbatim}
      USA       JPN       UKG       FRN       ITA       CHN 
1.4801300 1.0478834 0.6140597 0.5919777 0.4839285 0.4468015 
\end{verbatim}
\begin{alltt}
\hlcom{# additive "importer" effects}
\hlkwd{head}\hlstd{(} \hlkwd{sort}\hlstd{(bhat,}\hlkwc{decreasing}\hlstd{=}\hlnum{TRUE}\hlstd{)  )}
\end{alltt}
\begin{verbatim}
      USA       JPN       UKG       FRN       ITA       NTH 
1.5628243 0.8433793 0.6683700 0.5849702 0.4712668 0.3628532 
\end{verbatim}
\end{kframe}
\end{knitrout}
We note that these simple estimates here are 
very close to, but not exactly 
the same as, the least squares/maximum likelihood estimates
(this is because of the undefined diagonal in the sociomatrix). 

While straightforward to implement, this classical ANOVA analysis ignores 
a fundamental characteristic of dyadic data: Each node
appears in the dataset as both a sender and a receiver of 
relations, or equivalently, the 
row and column labels of the data matrix refer to the same set of 
objects. 
In the context of the ANOVA model, 
this means that 
each node $i$ has two additive effects: a row effect 
$a_i$  and a column effect $b_i$.  
Often it is of interest to evaluate the extent to which 
these effects are correlated, for example, to 
evaluate if sociable nodes in the network are also 
popular. 
Additionally, 
each  (unordered) pair of nodes $i,j$ has two 
outcomes, $y_{i,j}$ and $y_{j,i}$.  
It is often the case that
$y_{i,j}$ and $y_{j,i}$ are correlated, as these 
two observations come from the same dyad. 

Correlations between the additive effects can be evaluated empirically  
simply by computing the sample covariance of the 
row means and column means, or 
alternatively, the $\hat a_i$'s and 
$\hat b_i$'s. 
Dyadic correlation can be evaluated by computing the
correlation between the  matrix of residuals from the ANOVA 
model and its transpose:

\begin{knitrout}\footnotesize
\definecolor{shadecolor}{rgb}{0.969, 0.969, 0.969}\color{fgcolor}\begin{kframe}
\begin{alltt}
\hlcom{#### ---- covariance and correlation between row and column effects}
\hlkwd{cov}\hlstd{(} \hlkwd{cbind}\hlstd{(ahat,bhat) )}
\end{alltt}
\begin{verbatim}
          ahat      bhat
ahat 0.2407563 0.2290788
bhat 0.2290788 0.2289489
\end{verbatim}
\begin{alltt}
\hlkwd{cor}\hlstd{( ahat, bhat)}
\end{alltt}
\begin{verbatim}
[1] 0.9757237
\end{verbatim}
\end{kframe}
\end{knitrout}

\begin{knitrout}\footnotesize
\definecolor{shadecolor}{rgb}{0.969, 0.969, 0.969}\color{fgcolor}\begin{kframe}
\begin{alltt}
\hlcom{#### ---- an estimate of dyadic covariance and correlation}
\hlstd{R} \hlkwb{<-} \hlstd{Y} \hlopt{-} \hlstd{( muhat} \hlopt{+} \hlkwd{outer}\hlstd{(ahat,bhat,}\hlstr{"+"}\hlstd{) )}
\hlkwd{cov}\hlstd{(} \hlkwd{cbind}\hlstd{(} \hlkwd{c}\hlstd{(R),}\hlkwd{c}\hlstd{(}\hlkwd{t}\hlstd{(R)) ),} \hlkwc{use}\hlstd{=}\hlstr{"complete"}\hlstd{)}
\end{alltt}
\begin{verbatim}
          [,1]      [,2]
[1,] 0.2212591 0.1900891
[2,] 0.1900891 0.2212591
\end{verbatim}
\begin{alltt}
\hlkwd{cor}\hlstd{(} \hlkwd{c}\hlstd{(R),}\hlkwd{c}\hlstd{(}\hlkwd{t}\hlstd{(R)),} \hlkwc{use}\hlstd{=}\hlstr{"complete"}\hlstd{)}
\end{alltt}
\begin{verbatim}
[1] 0.8591242
\end{verbatim}
\end{kframe}
\end{knitrout}

\begin{figure}
\begin{knitrout}\footnotesize
\definecolor{shadecolor}{rgb}{0.969, 0.969, 0.969}\color{fgcolor}

{\centering \includegraphics[width=6in]{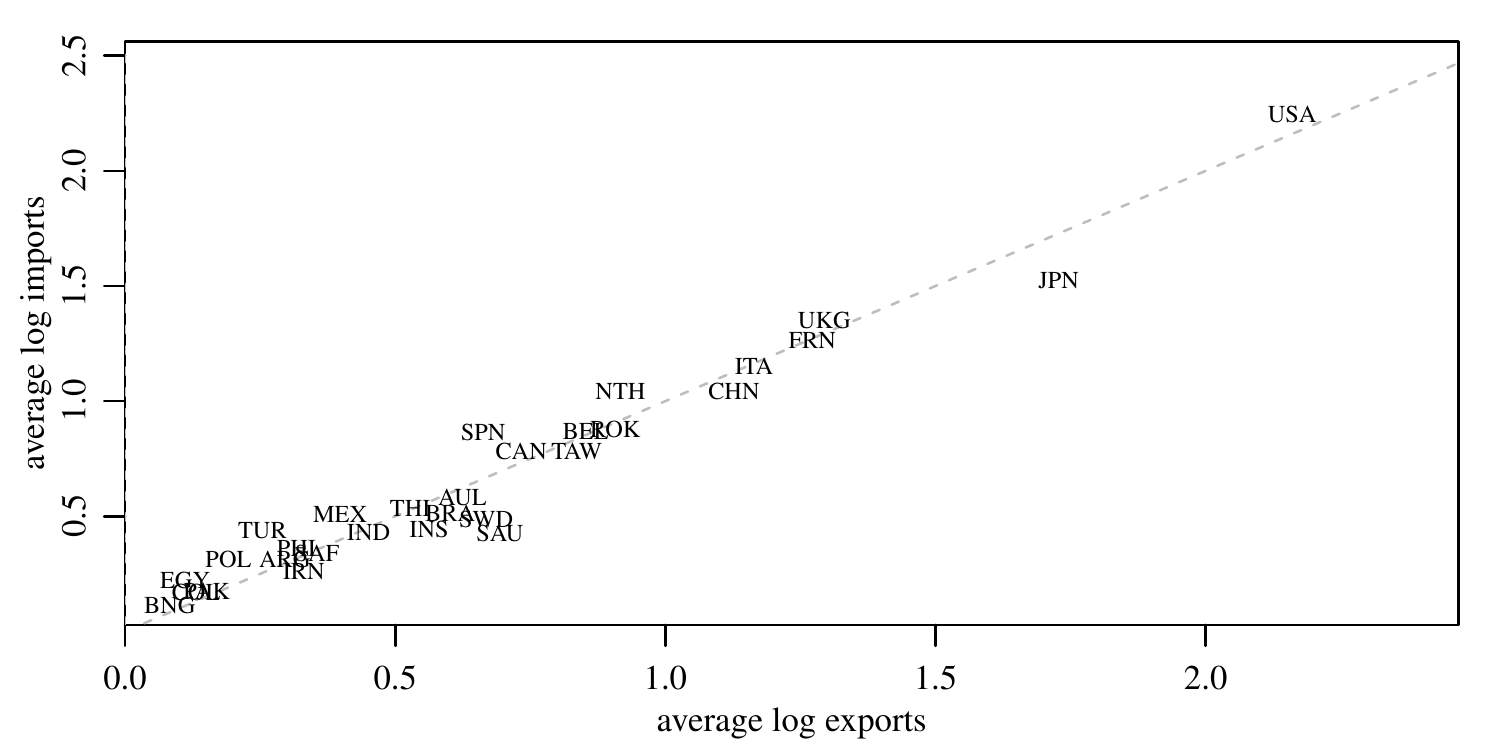} 

}

\end{knitrout}
\caption{Scatterplot of country-level average imports versus exports.} 
\label{fig:trade_rmvcm}
\end{figure}

As shown by these calculations and in Figure \ref{fig:trade_rmvcm}, 
country-level export and import volumes are highly correlated, 
as are the export and import volumes within country pairs. 
A seminal model for analyzing such 
within-node and within-dyad 
 dependence is 
the \emph{social relations model}, or SRM \citep{warner_kenny_stoto_1979}, 
a type of ANOVA decomposition that describes variability among the
 entries of the sociomatrix $\bl Y$ in terms of within-row, 
within-column and within-dyad variability.  
A normal random-effects version of the SRM has been studied by 
\citet{wong_1982} and  \citet{li_loken_2002}, among others, and 
takes the following form:
\begin{align} 
y_{i,j} & = \mu+ a_i + b_j + \epsilon_{i,j} \label{eqn:srm} \\
 \{ (a_1,b_1) ,\ldots, (a_n,b_n) \} &\sim  \text{i.i.d.} \ N(0,\Sigma_{ab})\nonumber  \\
 \{ (\epsilon_{i,j},\epsilon_{j,i}) : i\neq j \} &\sim  \text{i.i.d.} \ N(0,\Sigma_{e}) ,  \nonumber
\end{align}
where 
\[ 
\Sigma_{ab} = \begin{pmatrix} \sigma^2_a & \sigma_{ab} \\ 
    \sigma_{ab} & \sigma_b^2 \end{pmatrix} \ \  \text{and} \ \
\Sigma_\epsilon = \sigma^2_\epsilon \begin{pmatrix}  1 & \rho \\ \rho & 1 \end{pmatrix}. 
\]

Note that conditional on the row effects $\{a_1,\ldots, a_n\}$, 
the mean in the $i$th row of $\bl Y$ is given by $\mu+a_i$, 
and the variability of these row-specific  means is given by $\sigma^2_a$. 
In this way, the row effects represent across-row heterogeneity in the 
sociomatrix, and $\sigma_a^2$ is a single-number summary of this 
heterogeneity. 
Similarly, the column effects $\{b_1,\ldots, b_n\}$ 
represent heterogeneity in the column means, and 
$\sigma^2_b$ summarizes this heterogeneity. 
The covariance $\sigma_{ab}$ 
describes the linear association 
between 
these row and column effects, or equivalently, 
the association between the row means and column means of the sociomatrix. 
Additional variability across dyads is described by $\sigma^2_\epsilon$, 
and within dyad correlation  (beyond that 
  described by $\sigma_{ab}$) is captured by $\rho$. 
More precisely, straightforward calculations show that
under this random effects model, 
\begin{align} 
\Var{ y_{i,j} } &= \sigma^2_a + 2 \sigma_{ab} + \sigma^2_b + \sigma_\epsilon^2 
  & \text{(across-dyad variance)}  \label{eqn:srmcov} \\
\Cov{ y_{i,j} ,y_{i,k} } & =  \sigma_a^2  & \text{(within-row covariance)} \nonumber  \\ 
\Cov{ y_{i,j} ,y_{k,j} } & =  \sigma_b^2 & \text{(within-column covariance)}  \nonumber  \\
 \Cov{ y_{i,j} ,y_{j,k} } & =  \sigma_{ab} & \text{(row-column covariance) }    \nonumber   \\
 \Cov{ y_{i,j} ,y_{j,i} } & =  2 \sigma_{ab} + \rho \sigma^2_e & \text{(row-column covariance plus reciprocity) },  \nonumber
\end{align}
with all other covariances between elements of $\bl Y$ being zero. 
We refer to this covariance model as the \emph{social relations covariance
model}. 

The {\tt amen} package provides model fitting and evaluation tools 
for the SRM via the default values of the  {\tt ame} command:
\begin{knitrout}\footnotesize
\definecolor{shadecolor}{rgb}{0.969, 0.969, 0.969}\color{fgcolor}\begin{kframe}
\begin{alltt}
\hlstd{fit_SRM}\hlkwb{<-}\hlkwd{ame}\hlstd{(Y)}
\end{alltt}
\end{kframe}\begin{figure}

{\centering \includegraphics[width=\maxwidth]{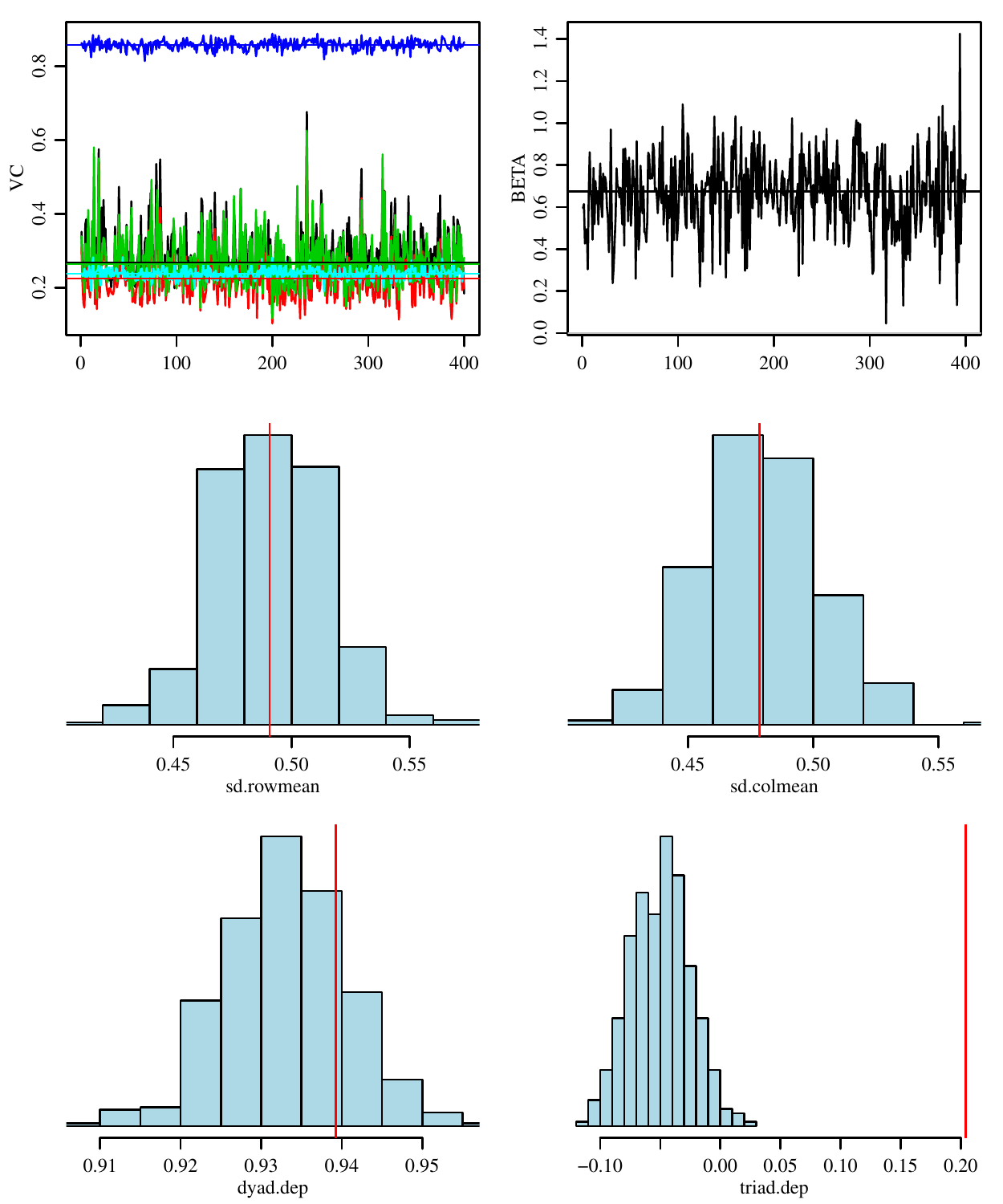} 

}

\caption[Default plots generated by the {\tt ame} command]{Default plots generated by the {\tt ame} command.}\label{fig:IR90s_fit_srm}
\end{figure}

\end{knitrout}

Running this command initiates an iterative Markov chain Monte Carlo  (MCMC) 
algorithm that provides Bayesian inference for the parameters in the  
SRM model. The progress of the algorithm is 
displayed via a sequence of plots,  
the last of 
which is shown in Figure \ref{fig:IR90s_fit_srm}. 
The top row gives
traceplots of the
parameter values
simulated from their posterior distribution, 
including covariance parameters on the
left and regression parameters on the right.
The covariance parameters include $\Sigma_{ab}$,
$\rho$, and $\sigma^2$, and are stored as
the list component {\tt VC} in the fitted object.
The only regression parameter for this  SRM model is the
intercept $\mu$, which is included by default for the Gaussian SRM.
The intercept, and any other regression parameters are
stored as {\tt BETA} in the fitted object.
We can compare these estimates obtained from {\tt amen}
to the estimates from the
ANOVA-style approach as follows:
\begin{knitrout}\footnotesize
\definecolor{shadecolor}{rgb}{0.969, 0.969, 0.969}\color{fgcolor}\begin{kframe}
\begin{alltt}
\hlstd{muhat}                                \hlcom{# empirical overall mean  }
\end{alltt}
\begin{verbatim}
[1] 0.680044
\end{verbatim}
\begin{alltt}
\hlkwd{mean}\hlstd{(fit_SRM}\hlopt{$}\hlstd{BETA)}                   \hlcom{# model-based estimate}
\end{alltt}
\begin{verbatim}
[1] 0.6616449
\end{verbatim}
\begin{alltt}
\hlkwd{cov}\hlstd{(} \hlkwd{cbind}\hlstd{(ahat,bhat) )}              \hlcom{# empirical row/column mean covariance }
\end{alltt}
\begin{verbatim}
          ahat      bhat
ahat 0.2407563 0.2290788
bhat 0.2290788 0.2289489
\end{verbatim}
\begin{alltt}
\hlkwd{apply}\hlstd{(fit_SRM}\hlopt{$}\hlstd{VC[,}\hlnum{1}\hlopt{:}\hlnum{3}\hlstd{],}\hlnum{2}\hlstd{,mean)}       \hlcom{# model-based estimate    }
\end{alltt}
\begin{verbatim}
       va       cab        vb 
0.2811301 0.2368096 0.2728049 
\end{verbatim}
\begin{alltt}
\hlkwd{cor}\hlstd{(} \hlkwd{c}\hlstd{(R),} \hlkwd{c}\hlstd{(}\hlkwd{t}\hlstd{(R)) ,} \hlkwc{use}\hlstd{=}\hlstr{"complete"}\hlstd{)} \hlcom{# empirical residual dyadic correlation}
\end{alltt}
\begin{verbatim}
[1] 0.8591242
\end{verbatim}
\begin{alltt}
\hlkwd{mean}\hlstd{(fit_SRM}\hlopt{$}\hlstd{VC[,}\hlnum{4}\hlstd{])}                 \hlcom{# model-based estimate}
\end{alltt}
\begin{verbatim}
[1] 0.857584
\end{verbatim}
\end{kframe}
\end{knitrout}
Posterior mean estimates of the row and column effects can be accessed
from {\tt fit\_SRM\$APM}  and {\tt fit\_SRM\$BPM}, respectively.
These estimates are plotted in Figure \ref{fig:bvols}, against the corresponding
ANOVA estimates. 
\begin{figure}
\begin{knitrout}\footnotesize
\definecolor{shadecolor}{rgb}{0.969, 0.969, 0.969}\color{fgcolor}

{\centering \includegraphics[width=5.2in]{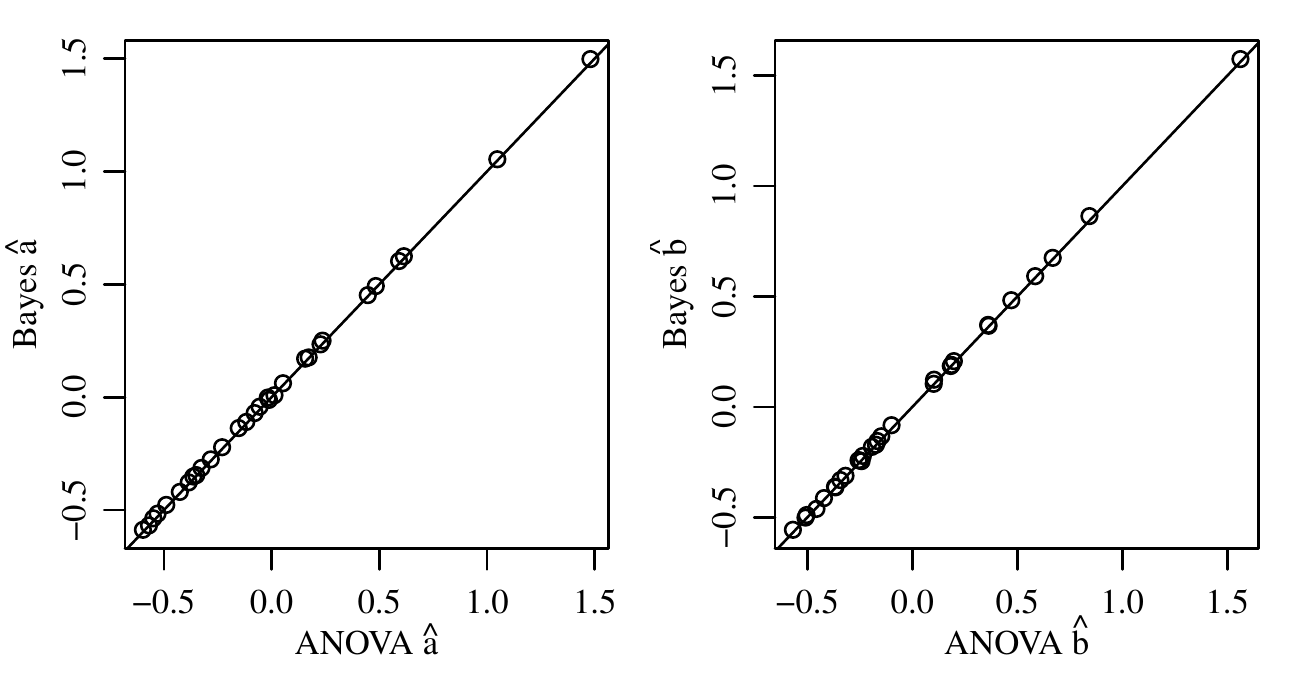} 

}

\end{knitrout}
\caption{Bayes versus least squares parameter estimates.} 
\label{fig:bvols}
\end{figure}

The second two rows of Figure  \ref{fig:IR90s_fit_srm}
 give posterior predictive
goodness of fit summaries for four network statistics:
(1) the empirical standard deviation of the row means;
(2) the empirical standard deviation of the column means;
(3) the empirical within-dyad  correlation;
(4)  a normalized measure of triadic dependence.
Details on how these are computed can be obtained by examining the
{\tt gofstats} function of the {\tt amen} package. 
The blue histograms in the figure represent values of
{\tt gofstats(Ysim)}, where {\tt Ysim} is simulated from the
posterior predictive distribution. These histograms
should be compared to the observed statistics {\tt gofstats(Y)},
which for these data are
0.491, 0.478, 0.939 
and 0.204, 
 given by vertical red lines in the figure.
Generally speaking,
large discrepancies between the posterior predictive distributions
(histograms) and the observed statistics (red lines) suggest
model lack of fit.
For these data, the model does well at representing the data 
with respect to the first three statistics, but shows a discrepancy with 
regard to 
the triadic dependence statistic. 
This is not too surprising, as the SRM only captures second-order 
dependencies (variances and covariances).

\subsection{Social relations regression modeling}
Often we wish to quantify the association between  
a particular dyadic variable and  
some other dyadic or nodal variables. 
Useful for such situations is a type of linear mixed effects 
model we refer to as the \emph{social relations regression model} (SRRM), 
which combines a linear regression model with the covariance 
structure of the SRM as follows:
\begin{equation}
 y_{i,j} = \beta_d^T \bl x_{d,i,j} + \beta_r^T \bl x_{r,i} +\beta_c^T \bl x_{c,j} +  a_i + b_j +  \epsilon_{i,j} ,  
\label{eqn:srrm}
\end{equation}
where $\bl x_{d,i,j}$ is a vector of characteristics of dyad $\{i,j\}$, 
    $\bl x_{r,i}$ is a vector of characteristics 
of node $i$ as  a sender, 
and $\bl x_{c,j}$ is a vector of characteristics of node $j$ 
as a receiver. 
We refer to $\bl x_{d,i,j}$, $\bl x_{r,i}$ and $\bl x_{c,i}$ 
as dyadic, row and column covariates, respectively. 
In many applications the row and column 
characteristics are the same so that $\bl x_{r,i} = \bl x_{c,i} = \bl x_i $, 
in which case they are simply referred to as nodal covariates. 
However, it can sometimes be useful to distinguish $\bl x_{r,i}$ from $\bl x_{c,i}$:
In the context 
of friendships among students, for example, it is conceivable that some 
characteristic  of a person 
(such as athletic or academic success) 
may affect their popularity (how much they are liked  by others), 
but not their sociability (how much they like others). 

We illustrate parameter estimation for the SRRM by fitting the model to the 
trade data. Nodal covariates include (log) population, (log) GDP, and polity, 
a measure of democracy. 
Dyadic covariates include the number the number of conflicts, (log) geographic 
distance between countries, the number of shared IGO memberships, and a polity 
interaction (the product of the nodal polity scores). 
\begin{knitrout}\footnotesize
\definecolor{shadecolor}{rgb}{0.969, 0.969, 0.969}\color{fgcolor}\begin{kframe}
\begin{alltt}
\hlcom{#### ---- nodal covariates}
\hlkwd{dimnames}\hlstd{(IR90s}\hlopt{$}\hlstd{nodevars)[[}\hlnum{2}\hlstd{]]}
\end{alltt}
\begin{verbatim}
[1] "pop"    "gdp"    "polity"
\end{verbatim}
\begin{alltt}
\hlstd{Xn}\hlkwb{<-}\hlstd{IR90s}\hlopt{$}\hlstd{nodevars[topgdp,]}
\hlstd{Xn[,}\hlnum{1}\hlopt{:}\hlnum{2}\hlstd{]}\hlkwb{<-}\hlkwd{log}\hlstd{(Xn[,}\hlnum{1}\hlopt{:}\hlnum{2}\hlstd{])}

\hlcom{#### ---- dyadic covariates}
\hlkwd{dimnames}\hlstd{(IR90s}\hlopt{$}\hlstd{dyadvars)[[}\hlnum{3}\hlstd{]]}
\end{alltt}
\begin{verbatim}
[1] "conflicts"   "exports"     "distance"    "shared_igos" "polity_int" 
\end{verbatim}
\begin{alltt}
\hlstd{Xd}\hlkwb{<-}\hlstd{IR90s}\hlopt{$}\hlstd{dyadvars[topgdp,topgdp,}\hlkwd{c}\hlstd{(}\hlnum{1}\hlstd{,}\hlnum{3}\hlstd{,}\hlnum{4}\hlstd{,}\hlnum{5}\hlstd{)]}
\hlstd{Xd[,,}\hlnum{3}\hlstd{]}\hlkwb{<-}\hlkwd{log}\hlstd{(Xd[,,}\hlnum{3}\hlstd{])}
\end{alltt}
\end{kframe}
\end{knitrout}
\noindent
Note that dyadic covariates are stored in an $n\times n \times p_d$ array, 
where $n$ is the number of nodes and $p_d$ is the number of dyadic covariates.

The SRRM can be fit 
by specifying the
covariates in the {\tt ame} function:
\begin{knitrout}\footnotesize
\definecolor{shadecolor}{rgb}{0.969, 0.969, 0.969}\color{fgcolor}\begin{kframe}
\begin{alltt}
\hlstd{fit_srrm}\hlkwb{<-}\hlkwd{ame}\hlstd{(Y,}\hlkwc{Xd}\hlstd{=Xd,}\hlkwc{Xr}\hlstd{=Xn,}\hlkwc{Xc}\hlstd{=Xn)}
\end{alltt}
\end{kframe}
\end{knitrout}
\noindent
Posterior mean estimates, standard deviations, nominal $z$-scores and $p$-values
may be obtained with the {\tt summary} command:
\begin{knitrout}\footnotesize
\definecolor{shadecolor}{rgb}{0.969, 0.969, 0.969}\color{fgcolor}\begin{kframe}
\begin{alltt}
\hlkwd{summary}\hlstd{(fit_srrm)}
\end{alltt}
\begin{verbatim}

Regression coefficients:
                  pmean   psd z-stat p-val
intercept        -6.407 1.255 -5.104 0.000
pop.row          -0.330 0.132 -2.502 0.012
gdp.row           0.567 0.151  3.764 0.000
polity.row       -0.015 0.020 -0.788 0.431
pop.col          -0.302 0.126 -2.388 0.017
gdp.col           0.537 0.147  3.647 0.000
polity.col       -0.006 0.019 -0.309 0.757
conflicts.dyad    0.076 0.042  1.822 0.068
distance.dyad    -0.041 0.007 -6.129 0.000
shared_igos.dyad  0.885 0.185  4.772 0.000
polity_int.dyad  -0.001 0.001 -1.668 0.095

Variance parameters:
    pmean   psd
va  0.264 0.104
cab 0.213 0.097
vb  0.250 0.098
rho 0.785 0.019
ve  0.157 0.010
\end{verbatim}
\end{kframe}
\end{knitrout}
The column {\tt z-stat} is obtained by dividing the posterior means by their 
posterior standard deviations, and each {\tt p-val} is the
the probability that a standard normal random variable 
exceeds the corresponding {\tt z-stat} in absolute value. 
Based on these calculations, there appears to be strong evidence for 
associations between countries' export and import levels with  both population and GDP. 
Additionally, there is evidence that geographic proximity and the number of 
shared IGOs are both positively associated with trade between country pairs.

It is instructive to compare these results to those that would be obtained 
under an ordinary linear regression model that assumes i.i.d.\ residual standard error. 
Such a model can be fit in the {\tt amen} package by opting 
to fit a model with no row variance, column variance or dyadic correlation:
\begin{knitrout}\footnotesize
\definecolor{shadecolor}{rgb}{0.969, 0.969, 0.969}\color{fgcolor}\begin{kframe}
\begin{alltt}
\hlstd{fit_rm}\hlkwb{<-}\hlkwd{ame}\hlstd{(Y,}\hlkwc{Xd}\hlstd{=Xd,}\hlkwc{Xr}\hlstd{=Xn,}\hlkwc{Xc}\hlstd{=Xn,}\hlkwc{rvar}\hlstd{=}\hlnum{FALSE}\hlstd{,}\hlkwc{cvar}\hlstd{=}\hlnum{FALSE}\hlstd{,}\hlkwc{dcor}\hlstd{=}\hlnum{FALSE}\hlstd{)}
\end{alltt}
\end{kframe}
\end{knitrout}

\begin{knitrout}\footnotesize
\definecolor{shadecolor}{rgb}{0.969, 0.969, 0.969}\color{fgcolor}\begin{kframe}
\begin{alltt}
\hlkwd{summary}\hlstd{(fit_rm)}
\end{alltt}
\begin{verbatim}

Regression coefficients:
                  pmean   psd  z-stat p-val
intercept        -4.417 0.170 -25.947 0.000
pop.row          -0.318 0.022 -14.621 0.000
gdp.row           0.664 0.024  27.417 0.000
polity.row       -0.007 0.005  -1.335 0.182
pop.col          -0.280 0.023 -12.328 0.000
gdp.col           0.622 0.024  25.590 0.000
polity.col        0.002 0.005   0.509 0.611
conflicts.dyad    0.238 0.057   4.152 0.000
distance.dyad    -0.053 0.004 -14.407 0.000
shared_igos.dyad -0.021 0.028  -0.739 0.460
polity_int.dyad   0.000 0.001   0.280 0.780

Variance parameters:
    pmean   psd
va  0.000 0.000
cab 0.000 0.000
vb  0.000 0.000
rho 0.000 0.000
ve  0.229 0.011
\end{verbatim}
\end{kframe}
\end{knitrout}

\begin{figure}
\begin{knitrout}\footnotesize
\definecolor{shadecolor}{rgb}{0.969, 0.969, 0.969}\color{fgcolor}

{\centering \includegraphics[width=6in]{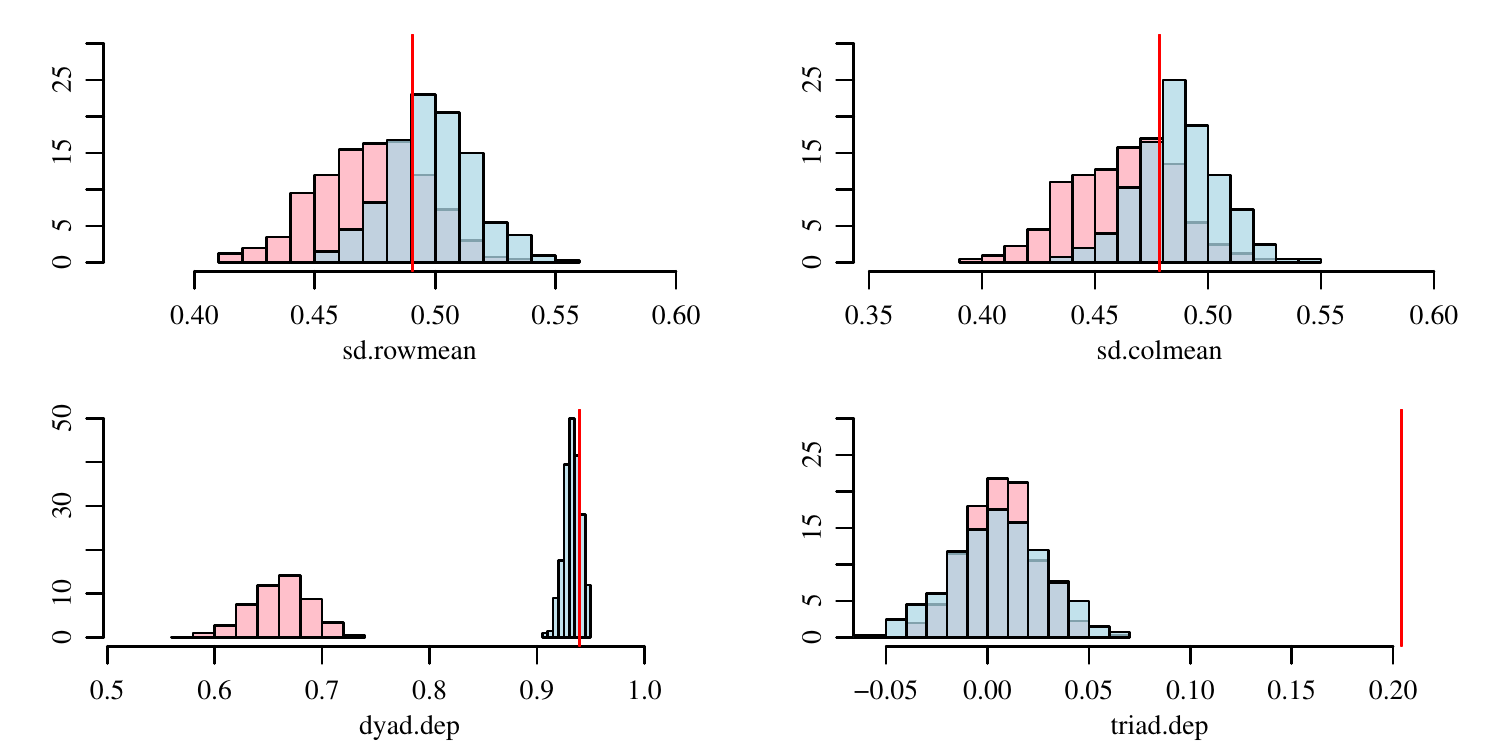} 

}

\end{knitrout}
\caption{Posterior predictive distributions of goodness of fit statistics for the 
ordinary regression model (pink) and the SRRM (blue).}
\label{fig:gof_srrm}
\end{figure}

The parameter standard deviations (i.e., standard errors) under this i.i.d.\  model are 
almost all smaller 
than those under the SRM fit. The explanation for this is that 
the i.i.d.\ model wrongly assumes independent observations, and thus 
overrepresents the precision of the parameter estimates. 
The inappropriateness of the i.i.d.\ model can be seen via 
the posterior predictive goodness of fit plots given in 
Figure \ref{fig:gof_srrm}. The plots show, in particular, that the data exhibit 
much more dyadic correlation than can be explained by the i.i.d.\ model. In 
contrast, the SRRM does not show such a discrepancy with regard to this statistic. 
However, both models fail to represent the amount of triadic dependence in the data, 
as shown in the fourth goodness of fit plot.

\subsection{Transitivity and stochastic equivalence via multiplicative 
effects}
It is often observed that the similarity of two nodes $i$ and $j$ in terms 
of their individual characteristics $\bl x_i$ and $\bl x_j$ 
is associated with the 
value of the relationship $y_{i,j}$ between them. 
For example, suppose  for each node $i$  that $x_i$ 
is the indicator that person $i$ 
is a member of a particular  group or organization. 
Then $x_{i}x_{j}$ is the indicator that $i$ and $j$ are 
co-members of this organization, and this fact may have 
some effect on their relationship $y_{i,j}$. 
A positive effect of $x_ix_j$ on $y_{i,j}$ is 
referred to as homophily, and a negative effect 
as anti-homophily. 
Measuring homophily on an observed 
characteristic can be done within the context of the SRRM 
by creating a dyadic covariate  $x_{d,i,j}$ 
from a nodal covariate $x_i$ through multiplication 
($x_{d,i,j} = x_i x_j$) or some other operation. 
Homophily on nodal characteristics can lead to certain types of 
patterns  often seen in network and dyadic data, 
sometimes referred to as transitivity, balance and clustering
\citep{hoff_2005a,hoff_2009c}. 
For example, in a binary network where people prefer to form 
ties to others who are similar to them,  
there tend to be a lot of 
 ``transitive triples,'' that is, triples of indices $i,j,k$ having a link 
between each pair. One explanation of this is that links 
from $i$ to $j$ and from $i$ to $k$ occur because $i$ is similar to 
both $j$ and to $k$. If this is the case, then 
$j$ and $k$ must also be somewhat similar, and so there 
is a high probability of a link between $j$ and $k$, which would form
a triangle of ties among nodes $i$, $j$ and $k$. Multiple linked 
triangles result in visual ``clusters'' in graphs of social networks. 

More generally, in the case of multiple sender and 
receiver covariates, 
we are interested in how a person with characteristics 
$\bl x_{r,i}$ relates to a person with characteristics 
$\bl x_{c,j}$. 
This can be evaluated in the SRRM by including a set of regression 
terms equivalent to 
$\bl x_{r,i}^T \bl B \bl x_{c,i}$. 
Although this term is multiplicative in the 
covariates, it is linear in the parameters, as 
\[ 
 \bl x_{r,i}^T \bl B \bl x_{c,i} = \sum_{k} \sum_l b_{k,l} x_{r,i,k} x_{c,j,l} 
\] 
and so the 
matrix of parameters may be estimated within the context of 
a linear regression model simply by including 
all products of the elements of $\bl x_{r,i}$ and $\bl x_{c,j}$ as 
dyadic covariates. In practice, 
if $\bl x_{r,i}$ and $\bl x_{c,j}$ are of the same length
(for example, if they are the same), 
then  it is 
common to take $\bl B$  to be a diagonal matrix, in which case  
\[ 
  \bl x_{r,i}^T \bl B \bl x_{c,i} = b_1 x_{r,i,1} x_{c,j,1}  + \cdots 
    + b_p x_{r,i,p} x_{c,j,p}. 
\] 
Such terms in the regression model can often account 
for network patterns such as transitivity and clustering, 
as described above. 
They can also account for another type of network pattern, 
known as stochastic equivalence, where it is 
observed that a group of nodes 
all relate to the other nodes (and each other) in a similar way. 
If such groups are related to the observed nodal covariates, then often 
the stochastic equivalence in the data  may be estimated and represented by 
these multiplicative regression terms. 

This can be seen to a limited degree in the trade data: 
Note that the number of shared IGOs and the polity interaction can both 
be viewed as dyadic covariates obtained by multiplication of 
nodal covariates. We can fit an SRRM without these effects as follows:
\begin{knitrout}\footnotesize
\definecolor{shadecolor}{rgb}{0.969, 0.969, 0.969}\color{fgcolor}\begin{kframe}
\begin{alltt}
\hlstd{fit_srrm0}\hlkwb{<-}\hlkwd{ame}\hlstd{(Y,Xd[,,}\hlnum{1}\hlopt{:}\hlnum{2}\hlstd{],Xn,Xn)}
\end{alltt}
\end{kframe}
\end{knitrout}

\begin{figure}
\begin{knitrout}\footnotesize
\definecolor{shadecolor}{rgb}{0.969, 0.969, 0.969}\color{fgcolor}
\includegraphics[width=\maxwidth]{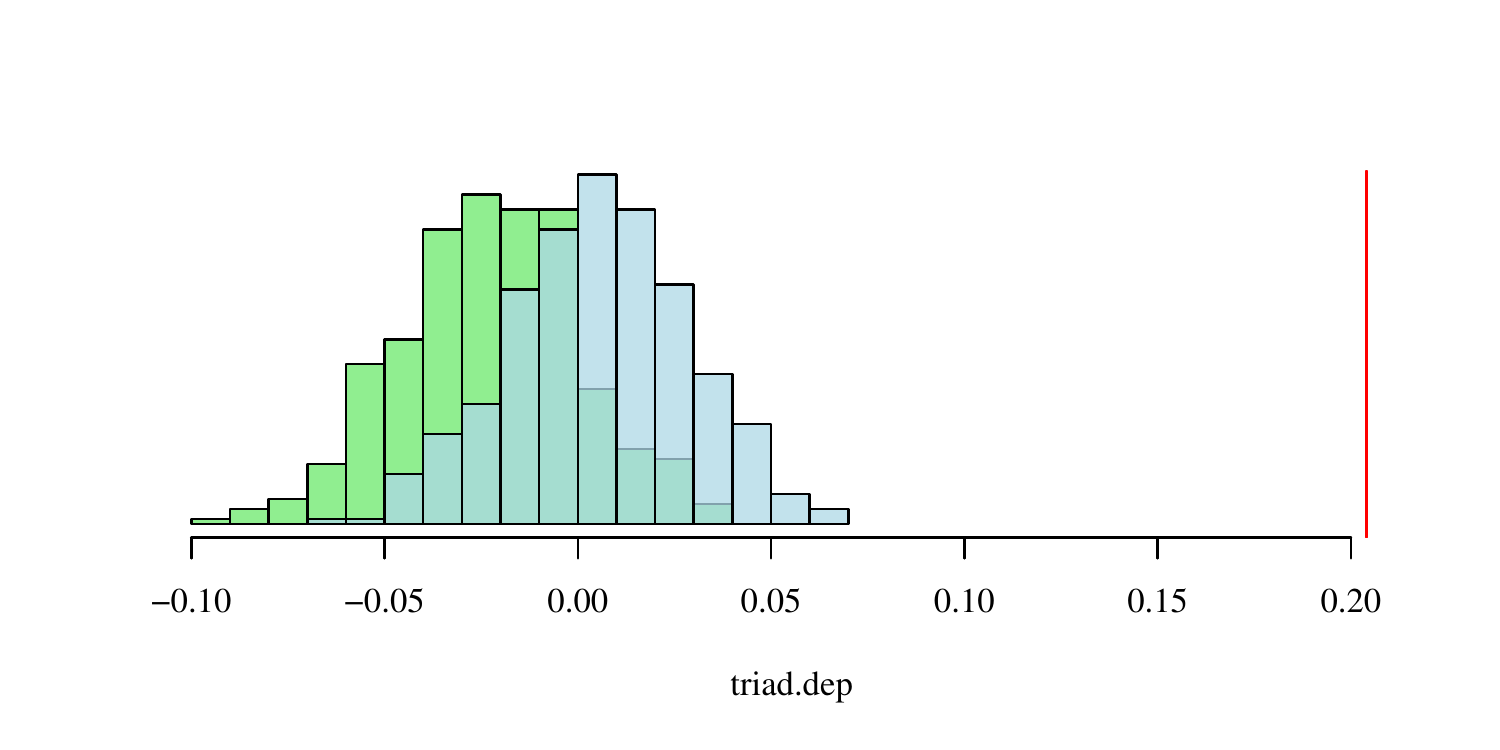} 

\end{knitrout}
\caption{Comparison of two  SRRMs
in terms of the triadic dependence statistic: 
with nodal
interaction effects (blue) and without (green). }
\label{fig:mdcomp}
\end{figure}

A comparison of the resulting posterior predictive distribution of the transitivity statistic
to that under the full SRRM (which included the multiplicative effects)
is given  in Figure \ref{fig:mdcomp}. 
The figure shows that, while both models do not fully represent the triadic dependence 
in the data, the model that includes the nodal interactions does slightly better. 
This raises the possibility that there may exist other nodal attributes, not given 
in the dataset, whose multiplicative interaction might help further describe the 
triadic dependence observed in the data. 
In such cases, 
it can be useful to include \emph{latent} nodal 
characteristics into the regression model, resulting in the following:
\begin{equation}
 y_{i,j} = \beta_d^T \bl x_{d,i,j} + \beta_r^T \bl x_{r,i} +\beta_c^T \bl x_{c,j} +  a_i + b_j +  \bl u_i^T \bl v_j  +  \epsilon_{i,j}. 
\label{eqn:ame}
\end{equation}
Here,  $\bl u_i$  is a vector of latent, unobserved factors or characteristics 
that describe node $i$'s behavior as a sender, and 
similarly $\bl v_j$  describes node $j$'s behavior as a receiver. 
In this model, the mean of $y_{i,j}$ 
depends on how ``similar'' $\bl u_i$ and $\bl v_j$ are (i.e., the extent to 
which the vectors point in the same direction) as well as the magnitudes of the 
vectors. 
Note also
 that  basic results from matrix algebra indicate that 
any type of network pattern that could be described by a regression 
term of the form  $\bl x_{r,i}^T \bl B \bl x_{c,j}$ can also 
be described by the multiplicative effects term $\bl u_i^T \bl v_j$.

We call a model of the form (\ref{eqn:ame}) an 
\emph{additive and multiplicative effects} model, or 
AME model for network and dyadic data. 
An AME model essentially combines two models for 
matrix-valued data: 
an \emph{additive main effects, multiplicative interaction} (AMMI) model
\citep{gollob_1968, bradu_gabriel_1974} - a class of 
models developed in the psychometric and agronomy literature; 
and the SRM covariance model that recognizes the dyadic 
aspect of the data. 
An AME model, like other latent factor models, requires the specification of the 
dimension of the latent factors. In the {\tt amen} package, this can be 
set with the option {\tt R} in the {\tt ame} command. 
The letter {\tt R} here stands for ``rank'':  If $\bl U$ and $\bl V$ are $n\times R$ matrices 
of the latent factors, then $\bl U\bl V^T$ has rank {\tt R}. 
For example, 
a rank-2 AME model  may be fit as follows:
\begin{knitrout}\footnotesize
\definecolor{shadecolor}{rgb}{0.969, 0.969, 0.969}\color{fgcolor}\begin{kframe}
\begin{alltt}
\hlstd{fit_ame2}\hlkwb{<-}\hlkwd{ame}\hlstd{(Y,Xd,Xn,Xn,}\hlkwc{R}\hlstd{=}\hlnum{2}\hlstd{)}
\end{alltt}
\end{kframe}\begin{figure}

{\centering \includegraphics[width=\maxwidth]{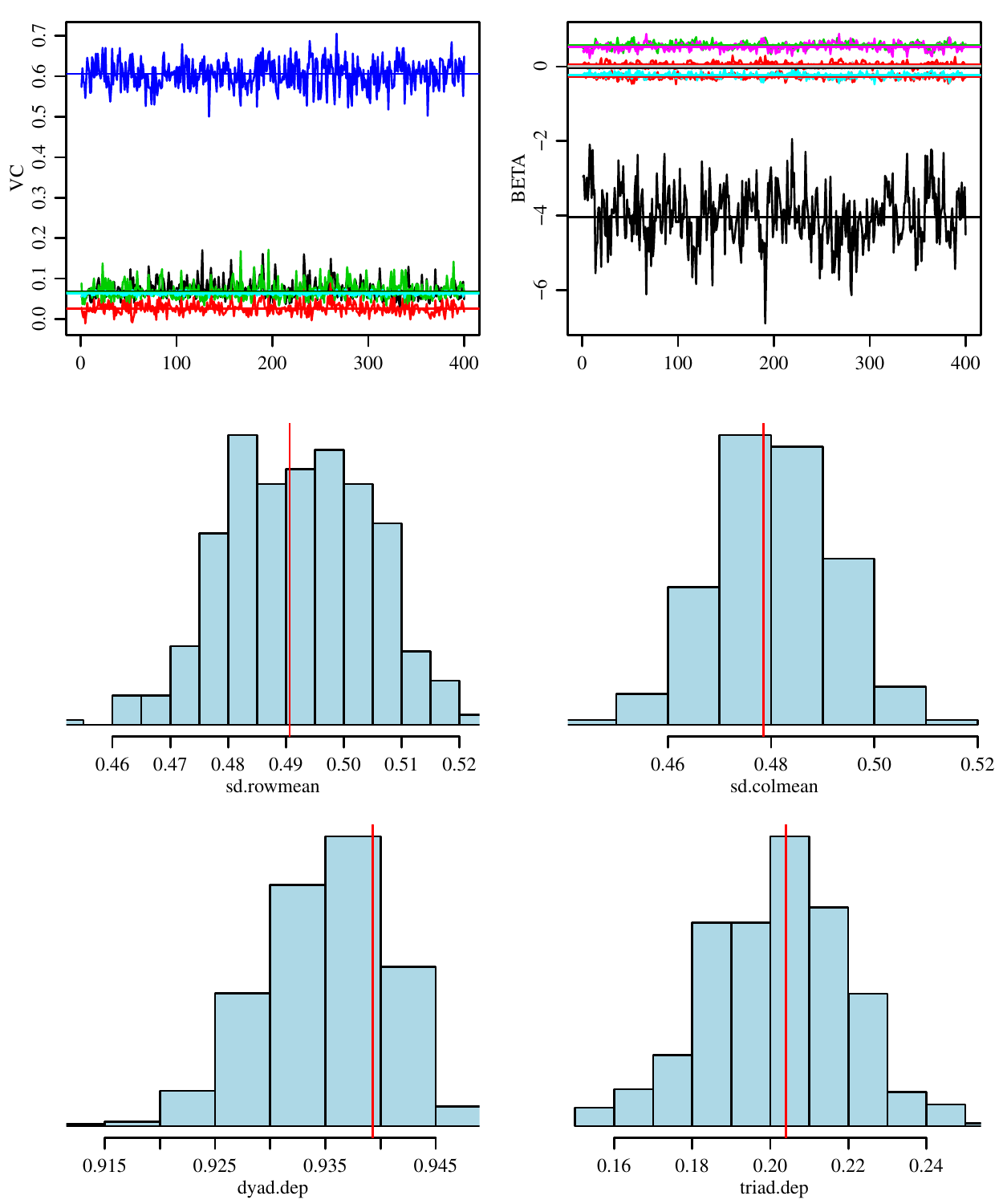} 

}

\caption[Diagnostic plots for the rank-2 AME model]{Diagnostic plots for the rank-2 AME model.}\label{fig:trade_ame2}
\end{figure}

\end{knitrout}
The diagnostic plots for this model  are given in Figure \ref{fig:trade_ame2}. 
Note that unlike all previous models considered, this model provides an adequate fit in terms 
of the triadic dependence statistic. 
The regression parameter estimates and their standard errors lead to more or
less similar conclusions as those from the SRRM, 
except that the number of shared IGOs
no longer has a large effect after controlling for the triadic dependence with the
latent factors.
\begin{knitrout}\footnotesize
\definecolor{shadecolor}{rgb}{0.969, 0.969, 0.969}\color{fgcolor}\begin{kframe}
\begin{alltt}
\hlkwd{summary}\hlstd{(fit_ame2)}
\end{alltt}
\begin{verbatim}

Regression coefficients:
                  pmean   psd z-stat p-val
intercept        -4.022 0.764 -5.263 0.000
pop.row          -0.277 0.069 -3.987 0.000
gdp.row           0.568 0.092  6.187 0.000
polity.row        0.000 0.010 -0.022 0.982
pop.col          -0.235 0.071 -3.290 0.001
gdp.col           0.525 0.099  5.315 0.000
polity.col        0.009 0.010  0.826 0.409
conflicts.dyad    0.018 0.036  0.513 0.608
distance.dyad    -0.039 0.004 -9.890 0.000
shared_igos.dyad  0.059 0.070  0.841 0.400
polity_int.dyad  -0.001 0.000 -2.273 0.023

Variance parameters:
    pmean   psd
va  0.072 0.022
cab 0.028 0.016
vb  0.070 0.021
rho 0.605 0.036
ve  0.063 0.004
\end{verbatim}
\end{kframe}
\end{knitrout}

\begin{figure}
\begin{knitrout}\footnotesize
\definecolor{shadecolor}{rgb}{0.969, 0.969, 0.969}\color{fgcolor}

{\centering \includegraphics[width=\maxwidth]{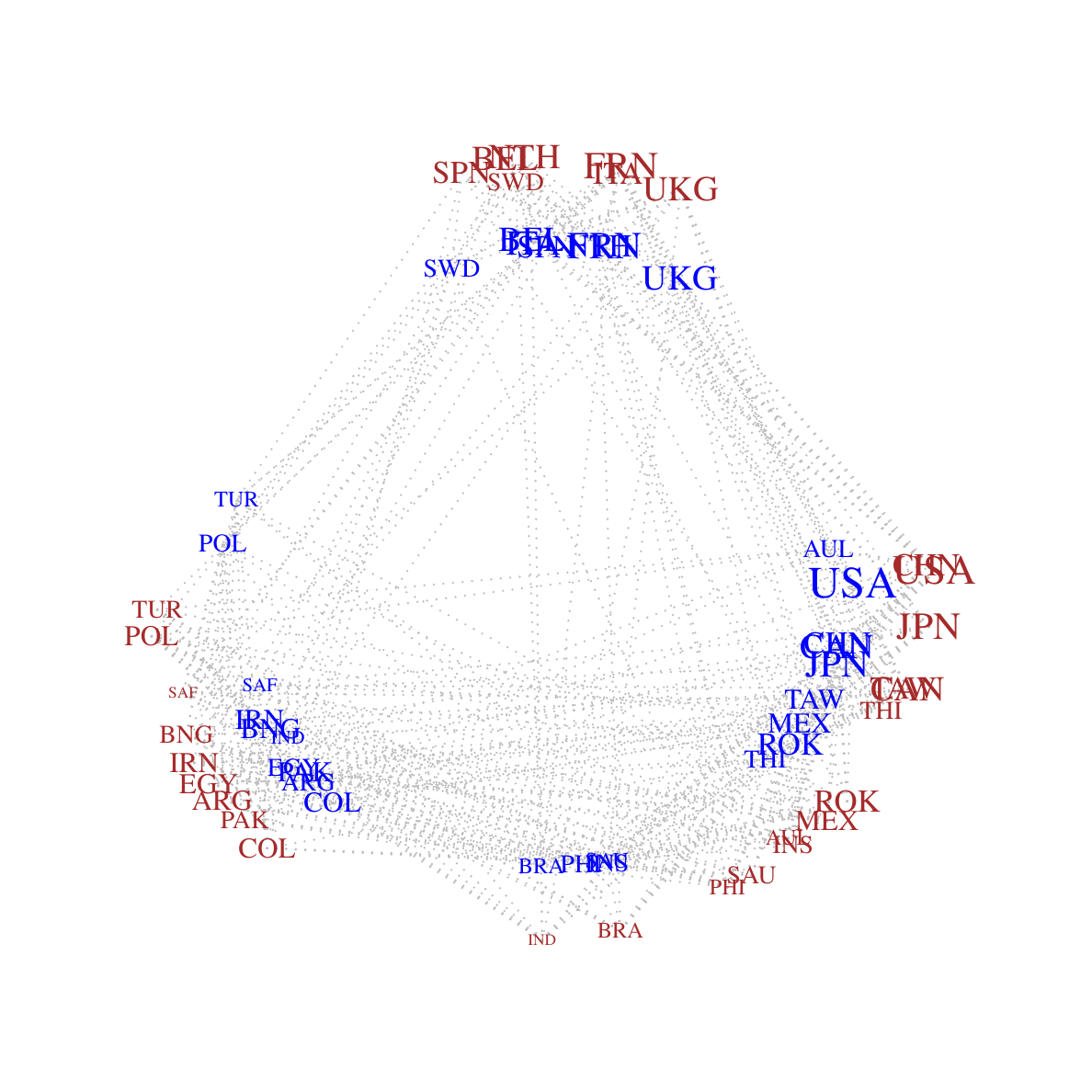} 

}

\end{knitrout}
\caption{Circle plot of estimated latent factors. Directions of 
$\hat {\bl u}_{i}$'s and $\hat {\bl v}_{i}$'s are given in red and 
blue, respectively, with the plotting size being a function of the 
magnitudes of the vectors.  Dashed lines between countries indicate 
greater than expected trade based on the regression terms 
and additive effects. }
\label{fig:trade_circplot}
\end{figure}

In some cases it is of interest to examine the estimated latent factors 
and compare them across nodes. Some ways to do this include clustering the 
latent factors or simply plotting them. The function {\tt circplot} in the 
{\tt amen} package provides a circle plot that can describe the estimated latent factors 
of a rank-2 model. A circle plot for the trade data is shown graphically in 
Figure \ref{fig:trade_circplot}. 
Such a figure can help identify groups of nodes that are similar to each 
other in terms of exporting and importing behavior, after controlling for 
regression and additive row and column effects. 
For example, the plot identifies the high trade volume between countries 
on the Pacific rim.


\section{AME models for ordinal data} 
Often we wish to analyze a dyadic outcome variable 
that is 
not well-represented by a normal model. 
In some cases, such as with the trade data, the variable of interest can be transformed 
so that the 
Gaussian 
AME model is reasonable.  In other cases, 
such as with binary, ordinal, discrete or sparse relations, 
 no such transformation is available.  
Examples of such data include measures of friendship that 
are binary (not friends/friends) or ordinal (dislike/neutral/like), 
discrete counts of conflictual events between countries, or the amount of time 
two people spend on the phone with each other  (which might be zero 
for most pairs in a population). 

In this section we describe extensions of the Gaussian AME model 
to accommodate ordinal dyadic data, where in what follows, 
ordinal means any outcome for which the possible values 
can be put in some meaningful order. This includes discrete outcomes 
(such as binary indicators or counts), ordered qualitative 
outcomes (such as low/medium/high, i.e.\  the
``traditional'' definition of ordinal), and even continuous outcomes. 
The extensions are based on latent variable 
representations of probit and ordinal probit regression models. 

\subsection{Example: Analysis of a binary outcome}
The simplest type of ordinal dyadic variable is a binary indicator of some 
type of relationship between $i$ and $j$, 
so that $y_{i,j} = $ 0 or 1 depending on whether the 
relationship  is absent or present, respectively. 
Such dyadic data, 
particularly data indicating 
social interactions or friendships, are often collectively called 
a \emph{social network}. For example, 
the {\tt amen} dataset {\tt lazegalaw} 
includes a social network of friendship ties between 71 members 
of a law firm, 
along with data on two other dyadic variables and several nodal 
variables. The friendship data are displayed as a graph in Figure 
\ref{fig:llaw_graph}, where the nodes are colored according to 
each lawyer's office location. 

\begin{knitrout}\footnotesize
\definecolor{shadecolor}{rgb}{0.969, 0.969, 0.969}\color{fgcolor}\begin{kframe}
\begin{alltt}
\hlkwd{data}\hlstd{(lazegalaw)}

\hlstd{Y}\hlkwb{<-}\hlstd{lazegalaw}\hlopt{$}\hlstd{Y[,,}\hlnum{2}\hlstd{]}
\hlstd{Xd}\hlkwb{<-}\hlstd{lazegalaw}\hlopt{$}\hlstd{Y[,,}\hlopt{-}\hlnum{2}\hlstd{]}
\hlstd{Xn}\hlkwb{<-}\hlstd{lazegalaw}\hlopt{$}\hlstd{X}

\hlkwd{dimnames}\hlstd{(Xd)[[}\hlnum{3}\hlstd{]]}
\end{alltt}
\begin{verbatim}
[1] "advice" "cowork"
\end{verbatim}
\begin{alltt}
\hlkwd{dimnames}\hlstd{(Xn)[[}\hlnum{2}\hlstd{]]}
\end{alltt}
\begin{verbatim}
[1] "status"    "female"    "office"    "seniority" "age"       "practice" 
[7] "school"   
\end{verbatim}
\begin{alltt}
\hlkwd{netplot}\hlstd{(lazegalaw}\hlopt{$}\hlstd{Y[,,}\hlnum{2}\hlstd{],}\hlkwc{ncol}\hlstd{=Xn[,}\hlnum{3}\hlstd{])}
\end{alltt}
\end{kframe}\begin{figure}

{\centering \includegraphics[width=\maxwidth]{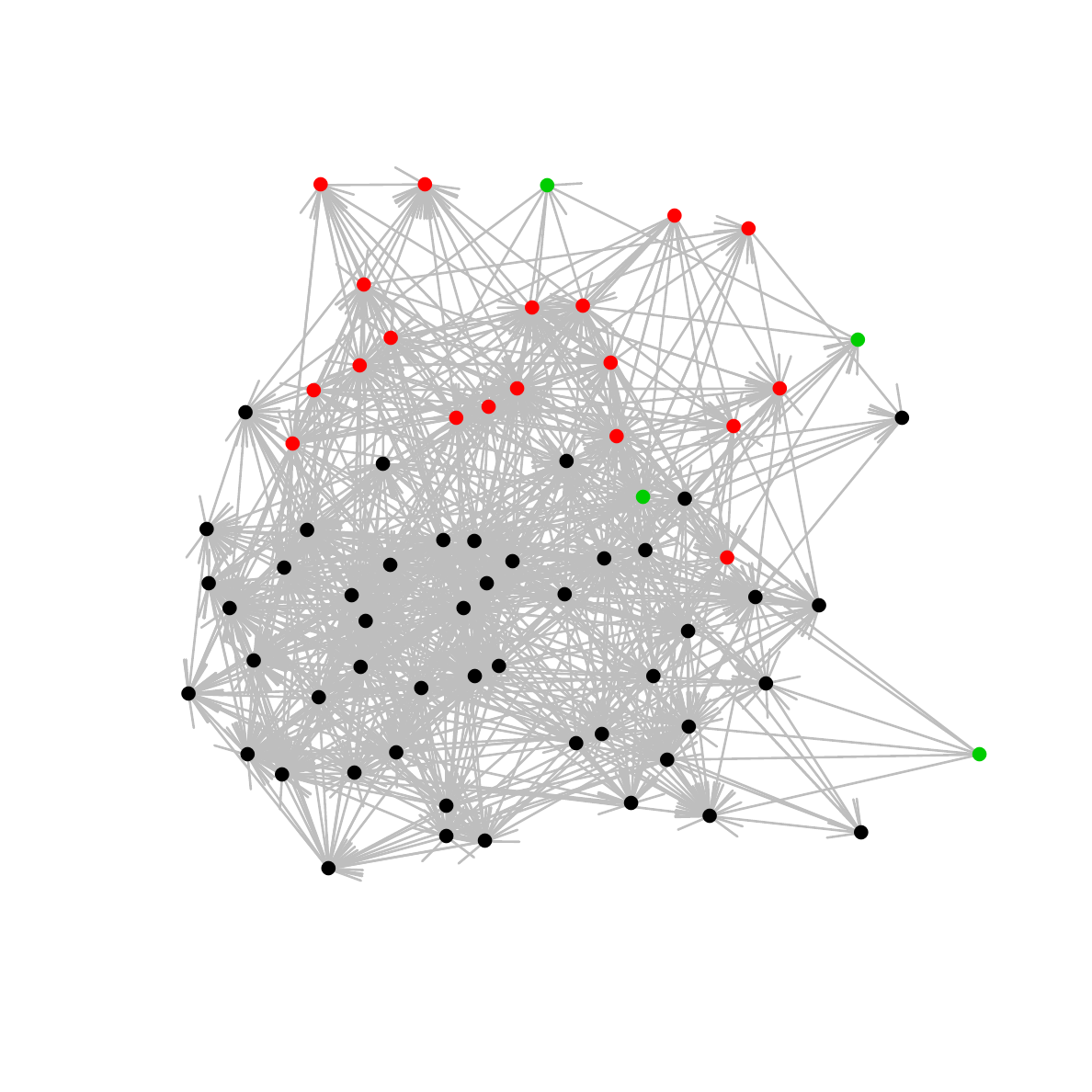} 

}

\caption[Graph of the friendship network between 71 lawyers]{Graph of the friendship network between 71 lawyers. Node colors represent at which of the three offices each lawyer works.}\label{fig:llaw_graph}
\end{figure}

\end{knitrout}

We first consider fitting a probit SRM model to 
these binary data, without including any explanatory covariates. 
This model can be written as
\begin{align}
z_{i,j} & = \mu + a_i + b_j + \epsilon_{i,j}  \label{eqn:psrm} \\
y_{i,j} & = 1(z_{i,j}>0), 
\end{align}
where the distributions of the random effects $a_i$, $b_j$, 
and $\epsilon_{i,j}$ follow the Gaussian SRM covariance 
model as described previously. 
This model expresses
the observed binary variable $y_{i,j}$ as the indicator that some 
continuous latent variable $z_{i,j}$ exceeds zero. 
Assuming the SRM for the sociomatrix $\bl Z=\{z_{i,j}\}$ of latent variables 
yields a model for the observed binary data that allows for 
within-row, within-column and within-dyad dependence. 
This model can be fit with the {\tt ame} command by specifying 
that the variable type is binary:
\begin{knitrout}\footnotesize
\definecolor{shadecolor}{rgb}{0.969, 0.969, 0.969}\color{fgcolor}\begin{kframe}
\begin{alltt}
\hlstd{fit_SRM}\hlkwb{<-}\hlkwd{ame}\hlstd{(Y,}\hlkwc{model}\hlstd{=}\hlstr{"bin"}\hlstd{)}
\end{alltt}
\end{kframe}
\end{knitrout}
It is instructive to compare the fit of this model to that provided 
by a reduced model that lacks the SRM terms:
\begin{knitrout}\footnotesize
\definecolor{shadecolor}{rgb}{0.969, 0.969, 0.969}\color{fgcolor}\begin{kframe}
\begin{alltt}
\hlstd{fit_SRG}\hlkwb{<-}\hlkwd{ame}\hlstd{(Y,}\hlkwc{model}\hlstd{=}\hlstr{"bin"}\hlstd{,}\hlkwc{rvar}\hlstd{=}\hlnum{FALSE}\hlstd{,}\hlkwc{cvar}\hlstd{=}\hlnum{FALSE}\hlstd{,}\hlkwc{dcor}\hlstd{=}\hlnum{FALSE}\hlstd{)}
\end{alltt}
\end{kframe}
\end{knitrout}
\noindent
This is a probit model that contains only an intercept, 
and so is equivalent to the simple random graph model (SRG). 
The fits of these two models in terms of the four 
goodness of fit statistics computed by 
{\tt gofstats} are compared in 
Figure 
\ref{fig:ll_fcomp}. 
As might be expected, the SRG fails in terms of all four statistics. 
In contrast, the SRM model provides a good fit in terms 
of the three statistics that represent  second-order dependence. 
Both models fail in terms of representing third-order 
dependence. 

\begin{figure}
\begin{knitrout}\footnotesize
\definecolor{shadecolor}{rgb}{0.969, 0.969, 0.969}\color{fgcolor}

{\centering \includegraphics[width=\maxwidth]{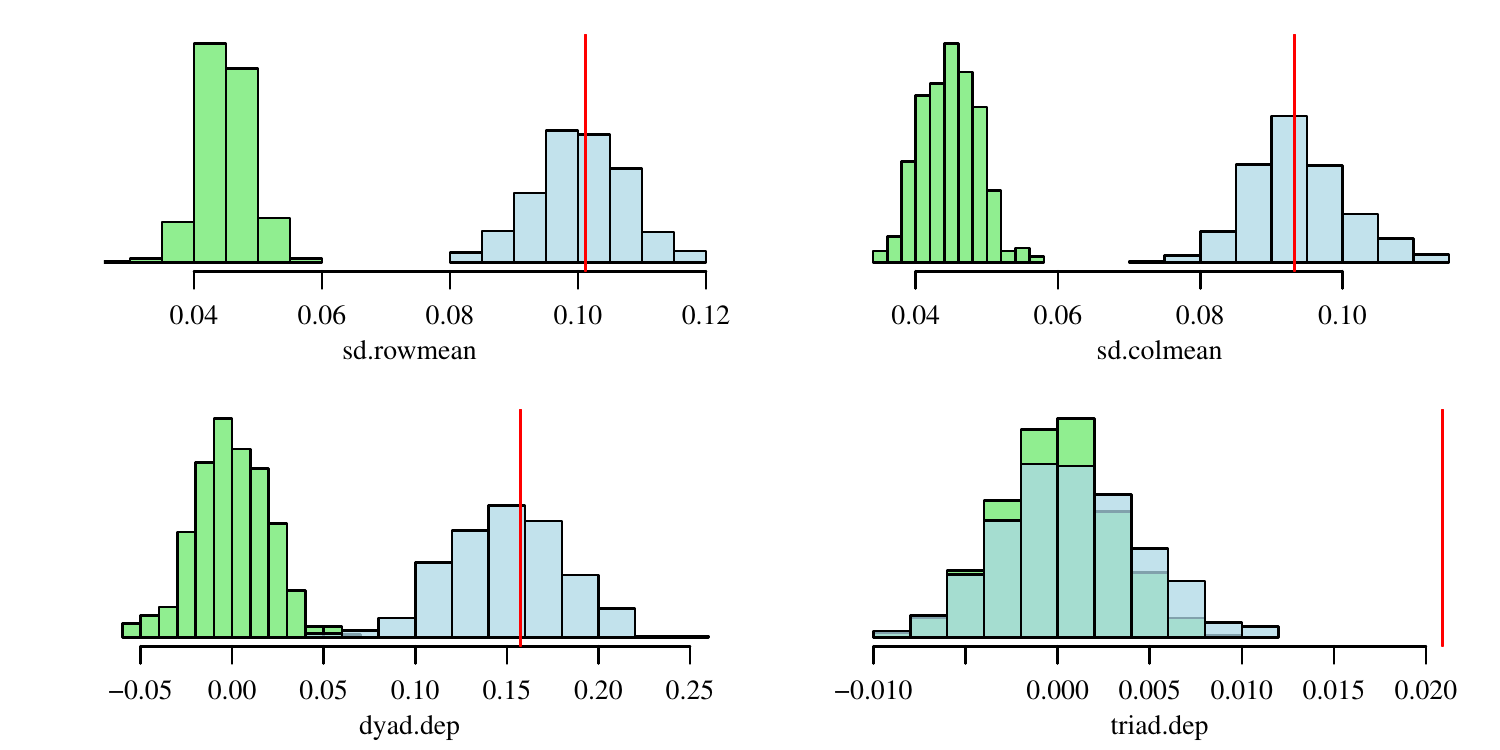} 

}

\end{knitrout}
\caption{Comparison of the SRM (blue) and  the SRG (green) for the 
Lazega law friendship network.} 
\label{fig:ll_fcomp}
\end{figure}

\begin{figure}
\begin{knitrout}\footnotesize
\definecolor{shadecolor}{rgb}{0.969, 0.969, 0.969}\color{fgcolor}

{\centering \includegraphics[width=\maxwidth]{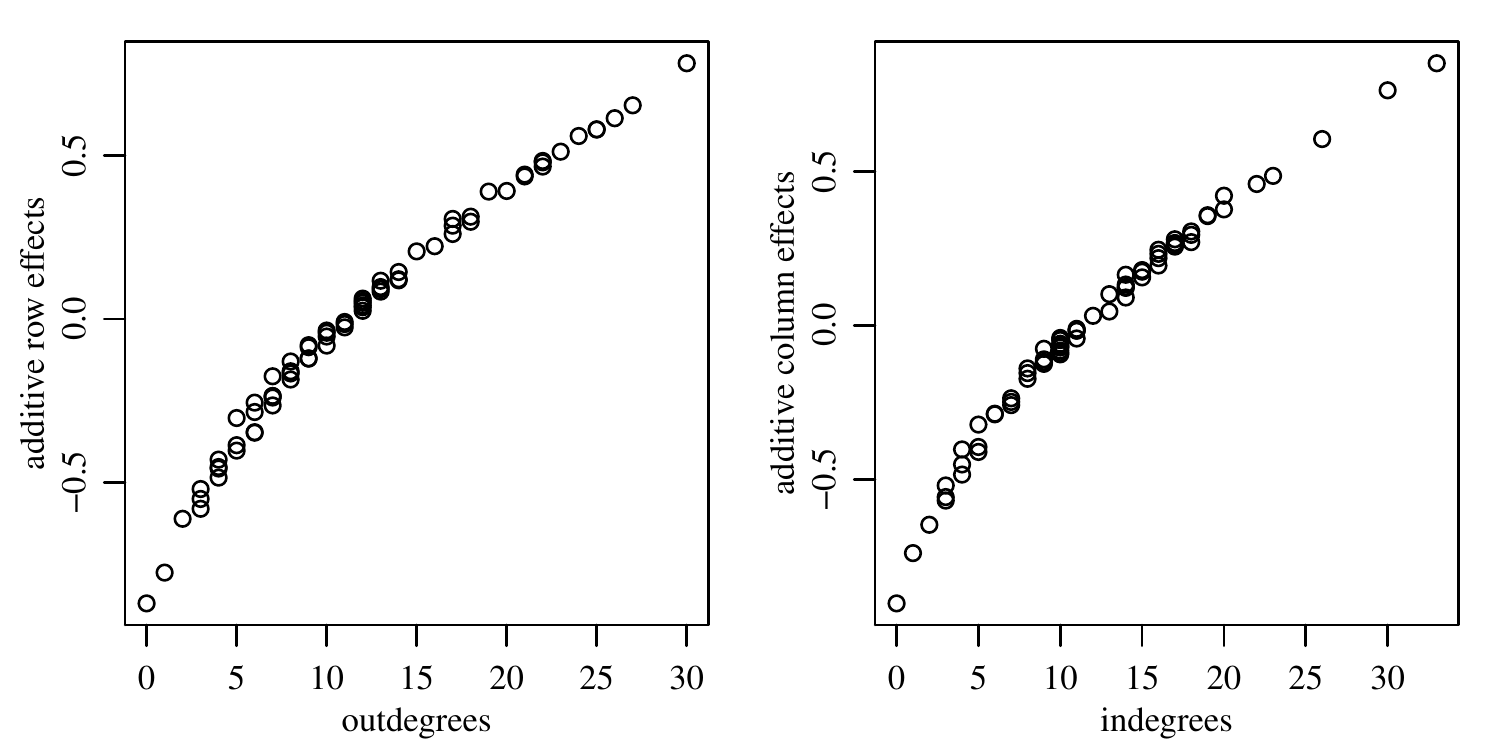} 

}

\end{knitrout}
\caption{Estimated row and column effects versus
outdegrees and indegrees.}
\label{fig:ll_degrees}
\end{figure}

A common empirical description of row and column heterogeneity 
in network data are 
the row and column sums, typically referred to as the 
\emph{outdegrees} and \emph{indegrees}. 
Based on the form of the model in (\ref{eqn:psrm}),
we might expect that the outdegrees and indegrees would be positively 
associated with the estimates of the $a_i$'s and $b_j$'s 
respectively. For example, the larger $a_i$ is, the larger the 
entries of $z_{i,j}$ for each $j$, thereby making more of the 
$y_{i,j}$'s equal to one rather than zero. 
This relationship between the degrees and the parameter estimates 
is illustrated in Figure 
\ref{fig:ll_degrees}. 
The figure does indeed show a strong positive association between these 
quantities, but note that the relationship is not 
strictly monotonic. The reason for this can be explained
by the fact that it is both the $a_i$ parameters \emph{and} 
the $b_j$ parameters that are used to describe nodal heterogeneity. 
For example, 
suppose two nodes have the same outdegree, but the 
first links to several nodes that have low indegrees, 
whereas a second node links to the same number of 
nodes but ones having high indegrees. 
The first node will have an $a_i$ estimate that is 
higher than that of the second, because the $b_j$'s 
of the nodes that the first links to will be lower 
than those of the nodes that the second links to.

We next consider a probit SRRM that includes the SRM terms
and 
linear regression effects for some 
 nodal and dyadic covariates. 
This model is formulated as 
in the SRM probit model, except that $z_{i,j}$ follows an 
SRRM rather than an SRM. 

\begin{knitrout}\footnotesize
\definecolor{shadecolor}{rgb}{0.969, 0.969, 0.969}\color{fgcolor}\begin{kframe}
\begin{alltt}
\hlstd{Xno}\hlkwb{<-}\hlstd{Xn[,}\hlkwd{c}\hlstd{(}\hlnum{1}\hlstd{,}\hlnum{2}\hlstd{,}\hlnum{4}\hlstd{,}\hlnum{5}\hlstd{,}\hlnum{6}\hlstd{)]}
\hlstd{fit_SRRM}\hlkwb{<-}\hlkwd{ame}\hlstd{(Y,} \hlkwc{Xd}\hlstd{=Xd,} \hlkwc{Xr}\hlstd{=Xno,} \hlkwc{Xc}\hlstd{=Xno,} \hlkwc{model}\hlstd{=}\hlstr{"bin"}\hlstd{)}
\end{alltt}
\end{kframe}
\end{knitrout}

\begin{knitrout}\footnotesize
\definecolor{shadecolor}{rgb}{0.969, 0.969, 0.969}\color{fgcolor}\begin{kframe}
\begin{alltt}
\hlkwd{summary}\hlstd{(fit_SRRM)}
\end{alltt}
\begin{verbatim}

Regression coefficients:
               pmean   psd z-stat p-val
intercept      0.882 0.659  1.338 0.181
status.row    -0.174 0.175 -0.992 0.321
female.row     0.007 0.143  0.051 0.959
seniority.row -0.008 0.012 -0.675 0.500
age.row       -0.016 0.009 -1.722 0.085
practice.row  -0.227 0.109 -2.089 0.037
status.col    -0.168 0.145 -1.154 0.248
female.col    -0.027 0.123 -0.219 0.827
seniority.col  0.012 0.011  1.056 0.291
age.col       -0.008 0.008 -1.064 0.287
practice.col  -0.286 0.110 -2.602 0.009
advice.dyad   -0.080 0.075 -1.069 0.285
cowork.dyad    1.281 0.063 20.313 0.000

Variance parameters:
    pmean   psd
va  0.170 0.037
cab 0.012 0.025
vb  0.134 0.032
rho 0.110 0.048
ve  1.000 0.000
\end{verbatim}
\end{kframe}
\end{knitrout}
There is not much evidence for effects of the 
nodal characteristics, at least in terms of  effects  
that appear linearly in the SRRM. 
Additionally, goodness-of-fit plots 
indicate lack of fit in terms of triadic dependence, 
as with the SRM model. 
Thus, we consider instead a model with the 
``non-significant'' regressors removed, and 
include a rank-3 multiplicative effect.

\begin{knitrout}\footnotesize
\definecolor{shadecolor}{rgb}{0.969, 0.969, 0.969}\color{fgcolor}\begin{kframe}
\begin{alltt}
\hlstd{fit_AME}\hlkwb{<-}\hlkwd{ame}\hlstd{(Y,} \hlkwc{Xd}\hlstd{=Xd[,,}\hlnum{2}\hlstd{],} \hlkwc{R}\hlstd{=}\hlnum{3}\hlstd{,} \hlkwc{model}\hlstd{=}\hlstr{"bin"}\hlstd{)}
\end{alltt}
\end{kframe}
\end{knitrout}

\begin{figure}
\begin{knitrout}\footnotesize
\definecolor{shadecolor}{rgb}{0.969, 0.969, 0.969}\color{fgcolor}

{\centering \includegraphics[width=\maxwidth]{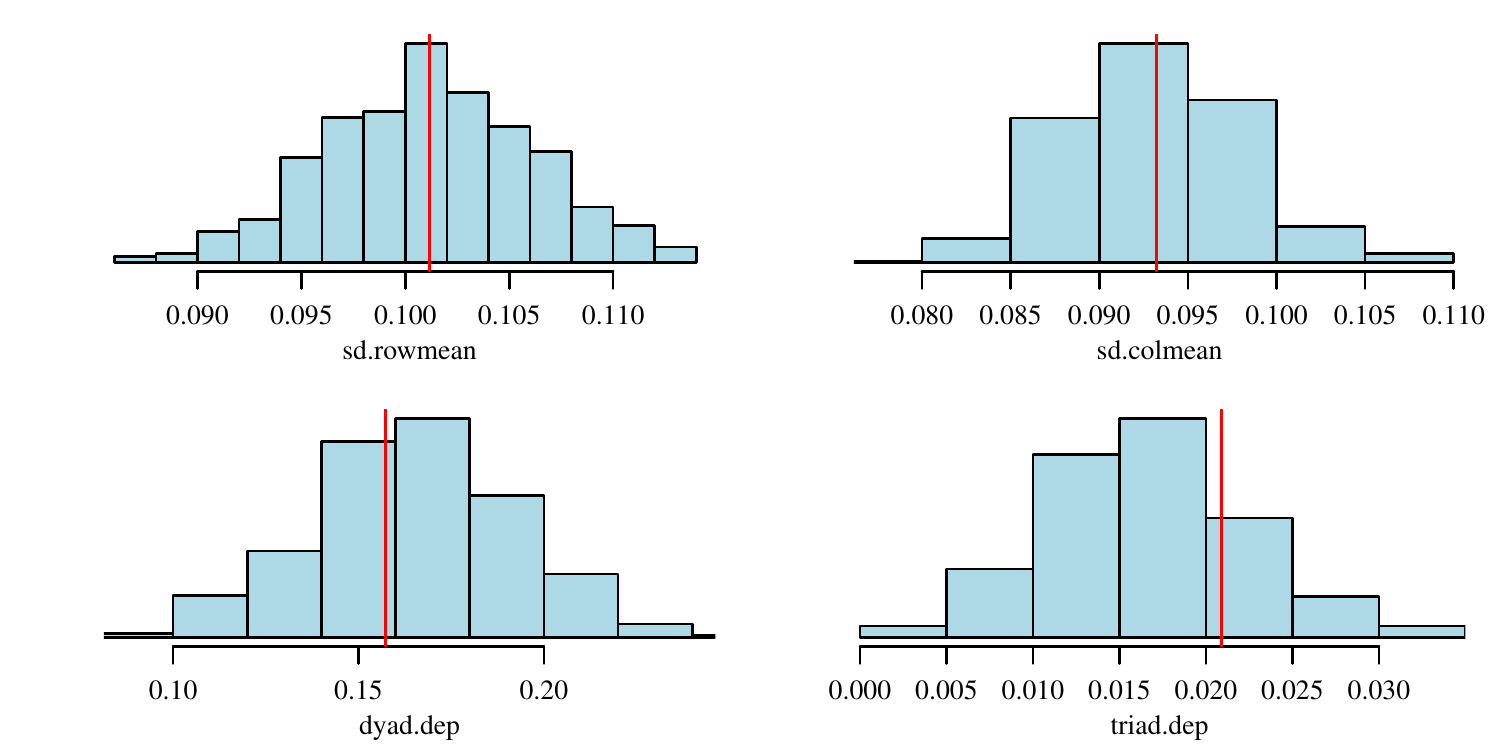} 

}

\end{knitrout}
\caption{ Checks of the fit of the rank-3 AME model to the 
Lazega law friendship network.}
\label{fig:ll_fame3}
\end{figure}

The goodness-of-fit plots in Figure \ref{fig:ll_fame3}
indicate no strong discrepancy between this model and 
the data in terms of these statistics. 
Inference then proceeds by examining
the estimates of regression effects, 
random effects and covariance parameters. 
Interpretation of the multiplicative effects 
can proceed by plotting them, looking for clusters, and 
identification of nodes with large effects. 
Additionally, it can be useful to look for associations between the 
multiplicative effects and any nodal characteristics available. 
For example, we can compute correlations between the
multiplicative effects $(\bl u_i,\bl v_i)$ and any 
numerical or ordinal nodal characteristics $\bl x_i$. 
Associations between multiplicative effects and categorical variables can be examined via plots. 

\begin{knitrout}\footnotesize
\definecolor{shadecolor}{rgb}{0.969, 0.969, 0.969}\color{fgcolor}\begin{kframe}
\begin{alltt}
\hlstd{U}\hlkwb{<-}\hlstd{fit_AME}\hlopt{$}\hlstd{U}
\hlstd{V}\hlkwb{<-}\hlstd{fit_AME}\hlopt{$}\hlstd{V}

\hlkwd{round}\hlstd{(}\hlkwd{cor}\hlstd{(U, Xno),}\hlnum{2}\hlstd{)}
\end{alltt}
\begin{verbatim}
     status female seniority   age practice
[1,]  -0.12  -0.02     -0.20 -0.25     0.00
[2,]  -0.34  -0.04      0.24  0.26     0.62
[3,]  -0.03   0.28     -0.01  0.12     0.20
\end{verbatim}
\begin{alltt}
\hlkwd{round}\hlstd{(}\hlkwd{cor}\hlstd{(V, Xno),}\hlnum{2}\hlstd{)}
\end{alltt}
\begin{verbatim}
     status female seniority  age practice
[1,]  -0.81  -0.32      0.71 0.62     0.24
[2,]  -0.04   0.15     -0.06 0.00     0.55
[3,]  -0.06   0.14      0.26 0.31     0.20
\end{verbatim}
\end{kframe}
\end{knitrout}

\begin{figure}
\begin{knitrout}\footnotesize
\definecolor{shadecolor}{rgb}{0.969, 0.969, 0.969}\color{fgcolor}
\includegraphics[width=\maxwidth]{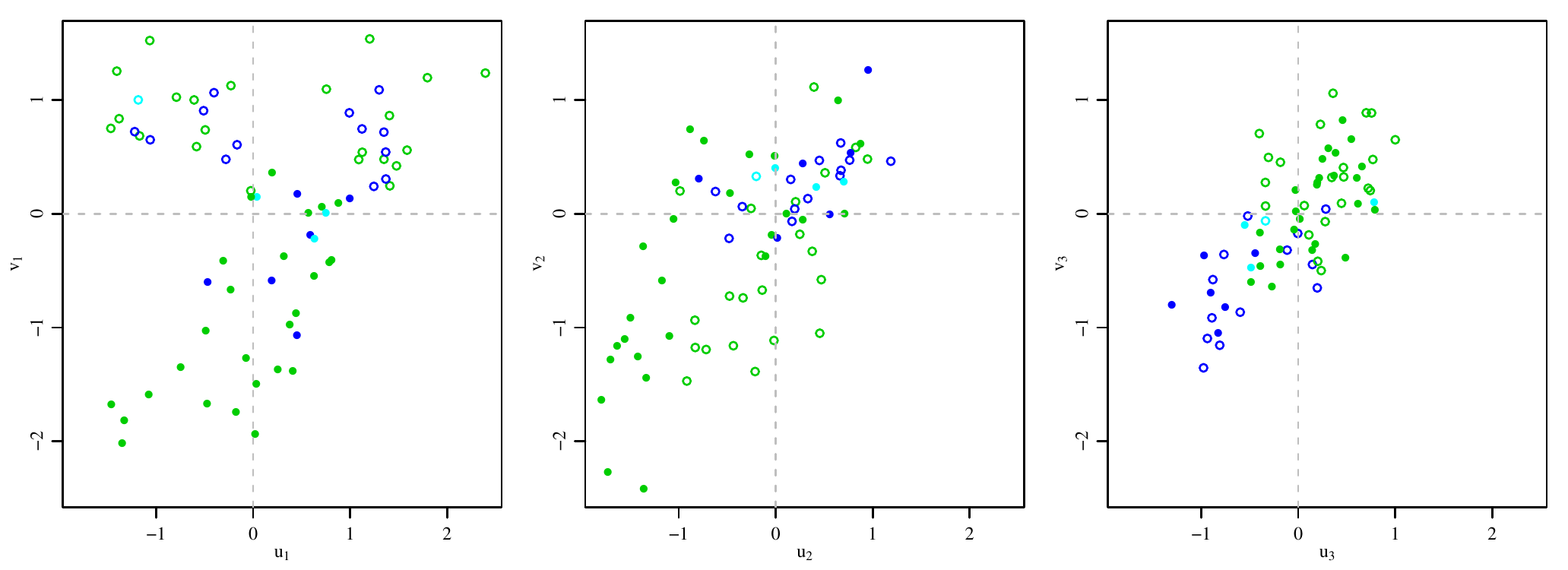} 

\end{knitrout}
\caption{Estimated latent factors plotted in terms of   
the nodal characteristics
  {\tt status} (partner=unfilled circle, associate=filled circle) and 
  {\tt office} (Boston=green, Hartford=blue, Providence=light blue).  }
\label{fig:ll_fplot}
\end{figure}

These correlations, and the plots in 
Figure \ref{fig:ll_fplot}, 
indicate that these nodal characteristics 
do play a role in network formation, although in 
a multiplicative rather than additive manner. 
If desired, one could use these results to 
construct multiplicative functions of these 
nodal attributes for inclusion into a SRRM, 
or possibly an 
AME model of lower rank.

\subsection{Example: Analysis of an ordinal outcome}
The probit AME model 
for binary data 
extends in a natural way to accommodate 
ordinal data with more than two levels. 
As with binary data, we model the 
sociomatrix $\bl Y =\{ y_{i,j} \}$ as being 
a function of a latent sociomatrix $\bl Z$ 
that follows a Gaussian AME model. Specifically, our model is 
\begin{align}
z_{i,j} & = \beta_d^T \bl x_{d,i,j} + \beta_r^T \bl x_{r,i} +\beta_c^T \bl x_{c,j} +  a_i + b_j +  \epsilon_{i,j} ,   \label{eqn:lame} \\
y_{i,j} & = g( z_{i,j} ) ,  \nonumber
\end{align}
where 
$g$ is some unknown non-decreasing function. 
The {\tt amen} package takes a 
semiparametric approach to this model, 
providing estimation and inference for the
parameters in the model (\ref{eqn:lame}) for $\bl Z$, 
but treating 
the function $g$ as a nuisance parameter. 
This is done using a variant of the 
\emph{extended 
rank likelihood} for ordinal data, described 
in \citet{hoff_2007a} and \citet{hoff_2008b}. 
While this approach is somewhat limiting
(as estimation of $g$ is not specifically provided), 
it simplifies some aspects of model 
specification and parameter estimation. 
In particular, the semiparametric approach 
allows for 
modeling of 
more general types of ordinal variables  $y_{i,j}$, such as
those that are
continuous, or those for which the number of 
levels is not pre-specified. 
However, we caution that the computation time required by the MCMC 
algorithm used by {\tt amen} is increasing in the number of 
levels of $y_{i,j}$.

We illustrate this model fitting procedure with an analysis of 
dominance relations between 28 female bighorn sheep, 
available via the {\tt sheep} dataset included with
{\tt amen}. 
The dyadic variable $y_{i,j}$ records the number of times 
sheep $i$ was observed dominating sheep $j$. 

\begin{knitrout}\footnotesize
\definecolor{shadecolor}{rgb}{0.969, 0.969, 0.969}\color{fgcolor}\begin{kframe}
\begin{alltt}
\hlkwd{data}\hlstd{(sheep)}

\hlstd{Y}\hlkwb{<-}\hlstd{sheep}\hlopt{$}\hlstd{dom}

\hlkwd{gofstats}\hlstd{(Y)}
\end{alltt}
\begin{verbatim}
 sd.rowmean  sd.colmean    dyad.dep   triad.dep 
 0.70037477  0.67209344 -0.19797403 -0.05826448 
\end{verbatim}
\end{kframe}
\end{knitrout}

Note that the dyadic dependence and triadic dependence 
statistics are negative. This makes sense in light of 
the nature of the variable: 
Heterogeneity among the sheep in terms of strength 
or assertiveness would lead to 
powerful sheep dominating but not being dominated by  others, 
thus leading to negative reciprocity. 
Additionally, under this scenario, if sheep $i$ dominated $j$, 
and $j$ dominated $k$, then it is unlikely that 
$k$ would be able to dominate $i$. Such a scenario would lead to negative 
triadic dependence.

Data on the ages of the sheep are also available. 
Plots of row and column means versus age are given in 
Figure \ref{fig:sheep}, and indicate some evidence of an 
age effect. Particularly, the number of times that 
a sheep is dominated is decreasing on average with age. 
We examine this effect more fully with an  ordinal probit regression, 
fitting a second-degree polynomial in the ages of the sheep:
\begin{figure}
\begin{knitrout}\footnotesize
\definecolor{shadecolor}{rgb}{0.969, 0.969, 0.969}\color{fgcolor}

{\centering \includegraphics[width=\maxwidth]{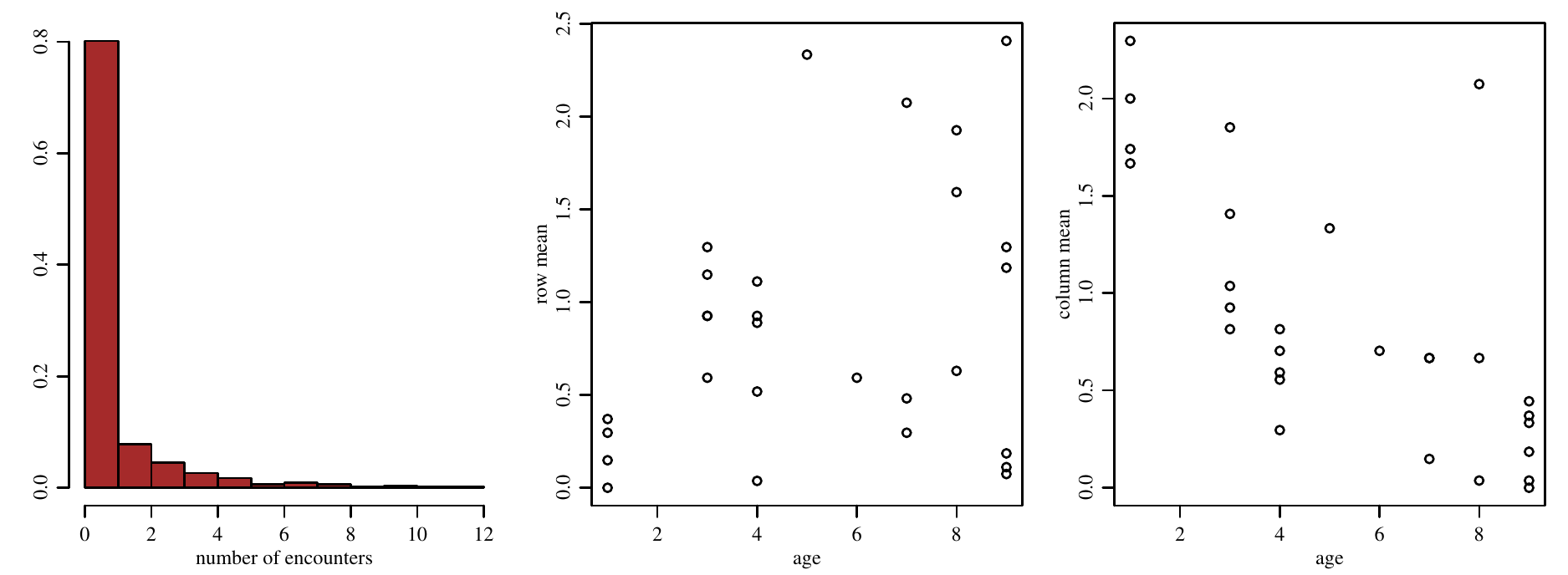} 

}

\end{knitrout}
\caption{Plots of the sheep dominance data. From left to right,
a histogram of the number of dominance encounters,
age versus row mean, and age versus column mean.}
\label{fig:sheep}
\end{figure}
\begin{knitrout}\footnotesize
\definecolor{shadecolor}{rgb}{0.969, 0.969, 0.969}\color{fgcolor}\begin{kframe}
\begin{alltt}
\hlstd{x}\hlkwb{<-}\hlstd{sheep}\hlopt{$}\hlstd{age} \hlopt{-} \hlkwd{mean}\hlstd{(sheep}\hlopt{$}\hlstd{age)}

\hlstd{Xd}\hlkwb{<-}\hlkwd{outer}\hlstd{(x,x)}

\hlstd{Xn}\hlkwb{<-}\hlkwd{cbind}\hlstd{(x,x}\hlopt{^}\hlnum{2}\hlstd{) ;} \hlkwd{colnames}\hlstd{(Xn)}\hlkwb{<-}\hlkwd{c}\hlstd{(}\hlstr{"age"}\hlstd{,}\hlstr{"age2"}\hlstd{)}

\hlstd{fit}\hlkwb{<-}\hlkwd{ame}\hlstd{(Y, Xd, Xn, Xn,} \hlkwc{model}\hlstd{=}\hlstr{"ord"}\hlstd{)}
\end{alltt}
\end{kframe}
\end{knitrout}

\begin{knitrout}\footnotesize
\definecolor{shadecolor}{rgb}{0.969, 0.969, 0.969}\color{fgcolor}\begin{kframe}
\begin{alltt}
\hlkwd{summary}\hlstd{(fit)}
\end{alltt}
\begin{verbatim}

Regression coefficients:
          pmean   psd z-stat p-val
age.row   0.158 0.051  3.101 0.002
age2.row -0.086 0.019 -4.454 0.000
age.col  -0.241 0.039 -6.136 0.000
age2.col -0.008 0.015 -0.561 0.575
.dyad     0.043 0.008  5.370 0.000

Variance parameters:
     pmean   psd
va   0.433 0.153
cab  0.039 0.073
vb   0.215 0.084
rho -0.399 0.091
ve   1.000 0.000
\end{verbatim}
\end{kframe}
\end{knitrout}
The results indicate evidence for a positive effect 
of age on dominance - older sheep are more likely 
to dominate and less likely to be dominated. 
The dyadic effect reflects some residual effect of homophily 
by age: A young sheep's dominance encounters are typically 
with other young sheep, and older sheep are more likely to 
be dominated by another older sheep than a younger sheep. 

Also note that the summary of the model fit does not include an intercept. 
This is because the intercept is not identifiable using the rank likelihood
approach used to obtain the parameter estimates. Specifically, 
an intercept term can be thought of as part of the transformation 
function $g$, which is being treated as a nuisance parameter.

\section{Censored and fixed rank nomination data}
Data on human social networks are often  
obtained by asking 
members of a study population to name 
a fixed number of people with whom they are friends, 
and possibly  to rank these friends in terms of their affinities to them. 
Such a survey method is called a \emph{fixed rank nomination} (FRN) 
scheme, and is commonly used in studies of institutions such as 
schools or businesses. For example, 
the National Longitudinal Study of Adolescent Health (AddHealth,
\citet{harris_2009}) asked middle and high-school students to nominate and 
rank up to five members of the same sex as friends, and five members 
of the opposite sex as friends. 

Data obtained from FRN schemes are similar to ordinal data, 
in that the ranks of a person's friends may be viewed as 
an ordinal response. However, FRN data are also censored
in a complicated way. 
Consider a study where people were asked to name and rank up to 
and including their top five friends. If person $i$ 
nominates five people but doesn't nominate person $j$, then $y_{i,j}$ is 
censored: The data cannot tell us whether $j$ is $i$'s sixth best 
friend, or whether $j$ is not liked by $i$ at all. 
On the other hand, if person $i$ nominates four people as friends 
but could have nominated five, then 
person $i$'s data are not censored -  the absence of a nomination 
by $i$ of $j$ indicates that $i$ does not consider $j$ a friend. 

A likelihood-based approach to modeling 
FRN data was developed in 
\citet{hoff_fosdick_volfovsky_stovel_2013}. 
Similar to the approach for ordinal dyadic data described above, 
this methodology treats the observed ranked outcomes $\bl Y$ 
as a function of an underlying continuous sociomatrix $\bl Z$ of affinities 
that 
is generated from an 
AME model. 
Letting $m$ be the maximum number of nominations allowed, and 
coding $y_{i,j} \in \{ m,m-1,\ldots, 1,0\} 
$ so that $y_{i,j}=m$ indicates that $j$ is $i$'s most liked friend, 
the FRN likelihood is derived from the following constraints that the 
observed ranks $\bl Y$ tell us about the underlying dyadic variables $\bl Z$:
\begin{eqnarray}
y_{i,j}> 0 &\Rightarrow & z_{i,j}>0  \label{eqn:porc}\\
y_{i,j}  > y_{i,k} & \Rightarrow &  z_{i,j} > z_{i,k} 
  \label{eqn:rnkc}\\ 
y_{i,j} = 0  \ \mbox{and} \  
 d_i < m &\Rightarrow & 
    z_{i,j}\leq 0  \label{eqn:degc}. 
\end{eqnarray}
Constraint (\ref{eqn:porc}) 
indicates that 
if $i$ ranks $j$, then $i$ has a positive relation with $j$ ($z_{i,j}>0$), 
and constraint (\ref{eqn:rnkc})  indicates that 
a higher rank corresponds to a more positive relation. 
Letting $d_i\in \{ 0,\ldots, m\}$ be the number of people 
that $i$ ranks, constraint (\ref{eqn:degc})  
indicates that if 
$i$ could have made additional friendship nominations 
but chose not to nominate $j$, 
they then must not consider $j$ a friend. 
On the other hand, if
$y_{i,j}=0$ but $d_i = m$ then person $i$'s unranked relationships
are censored, and so $z_{i,j}$ could be positive
even though
$y_{i,j}=0$.  In this case, all that is known about
$z_{i,j}$ is that it is less than $z_{i,k}$ for any person
$k$ that is ranked by $i$.

\subsection{Example: Analysis of fixed rank nomination data}
The {\tt amen} package implements a Bayesian model fitting 
algorithm based on the FRN likelihood. 
We illustrate its use with an analysis of  data from the classic 
study on relationships  between monks 
described in \citet{sampson_1969}, in which each monk was asked to rank 
up to three other monks in terms of a variety of relations. 
\begin{knitrout}\footnotesize
\definecolor{shadecolor}{rgb}{0.969, 0.969, 0.969}\color{fgcolor}\begin{kframe}
\begin{alltt}
\hlstd{Y}\hlkwb{<-}\hlstd{sampsonmonks[,,}\hlnum{3}\hlstd{]}

\hlkwd{apply}\hlstd{(Y}\hlopt{>}\hlnum{0}\hlstd{,}\hlnum{1}\hlstd{,sum,}\hlkwc{na.rm}\hlstd{=T)}
\end{alltt}
\begin{verbatim}
  ROMUL BONAVEN AMBROSE   BERTH   PETER   LOUIS  VICTOR    WINF    JOHN 
      3       3       4       3       3       3       3       3       3 
   GREG    HUGH    BONI    MARK  ALBERT   AMAND   BASIL   ELIAS    SIMP 
      4       3       3       3       3       3       3       3       3 
\end{verbatim}
\end{kframe}
\end{knitrout}
Notice that two of the monks didn't follow the survey instructions, and 
nominated more than three other monks. We treat the maximum 
number of nominations for these two monks as four. 
This can be done using the {\tt ame} fitting function, and 
specifying the FRN likelihood and the number of maximum nominations as follows:
\begin{knitrout}\footnotesize
\definecolor{shadecolor}{rgb}{0.969, 0.969, 0.969}\color{fgcolor}\begin{kframe}
\begin{alltt}
\hlstd{odmax}\hlkwb{<-}\hlkwd{rep}\hlstd{(}\hlnum{3}\hlstd{,}\hlkwd{nrow}\hlstd{(Y))}
\hlstd{odmax[} \hlkwd{apply}\hlstd{(Y}\hlopt{>}\hlnum{0}\hlstd{,}\hlnum{1}\hlstd{,sum,}\hlkwc{na.rm}\hlstd{=T)}\hlopt{>}\hlnum{3} \hlstd{]}\hlkwb{<-}\hlnum{4}

\hlstd{fit}\hlkwb{<-}\hlkwd{ame}\hlstd{(Y,}\hlkwc{R}\hlstd{=}\hlnum{2}\hlstd{,}\hlkwc{model}\hlstd{=}\hlstr{"frn"}\hlstd{,}\hlkwc{odmax}\hlstd{=odmax)}
\end{alltt}
\end{kframe}
\end{knitrout}

\begin{knitrout}\footnotesize
\definecolor{shadecolor}{rgb}{0.969, 0.969, 0.969}\color{fgcolor}\begin{kframe}
\begin{alltt}
\hlkwd{summary}\hlstd{(fit)}
\end{alltt}
\begin{verbatim}

Regression coefficients:
          pmean   psd z-stat p-val
intercept  0.64 0.725  0.883 0.377

Variance parameters:
    pmean   psd
va  0.557 0.768
cab 0.006 0.152
vb  0.249 0.185
rho 0.761 0.164
ve  1.000 0.000
\end{verbatim}
\end{kframe}
\end{knitrout}

\begin{figure}
\begin{knitrout}\footnotesize
\definecolor{shadecolor}{rgb}{0.969, 0.969, 0.969}\color{fgcolor}

{\centering \includegraphics[width=\maxwidth]{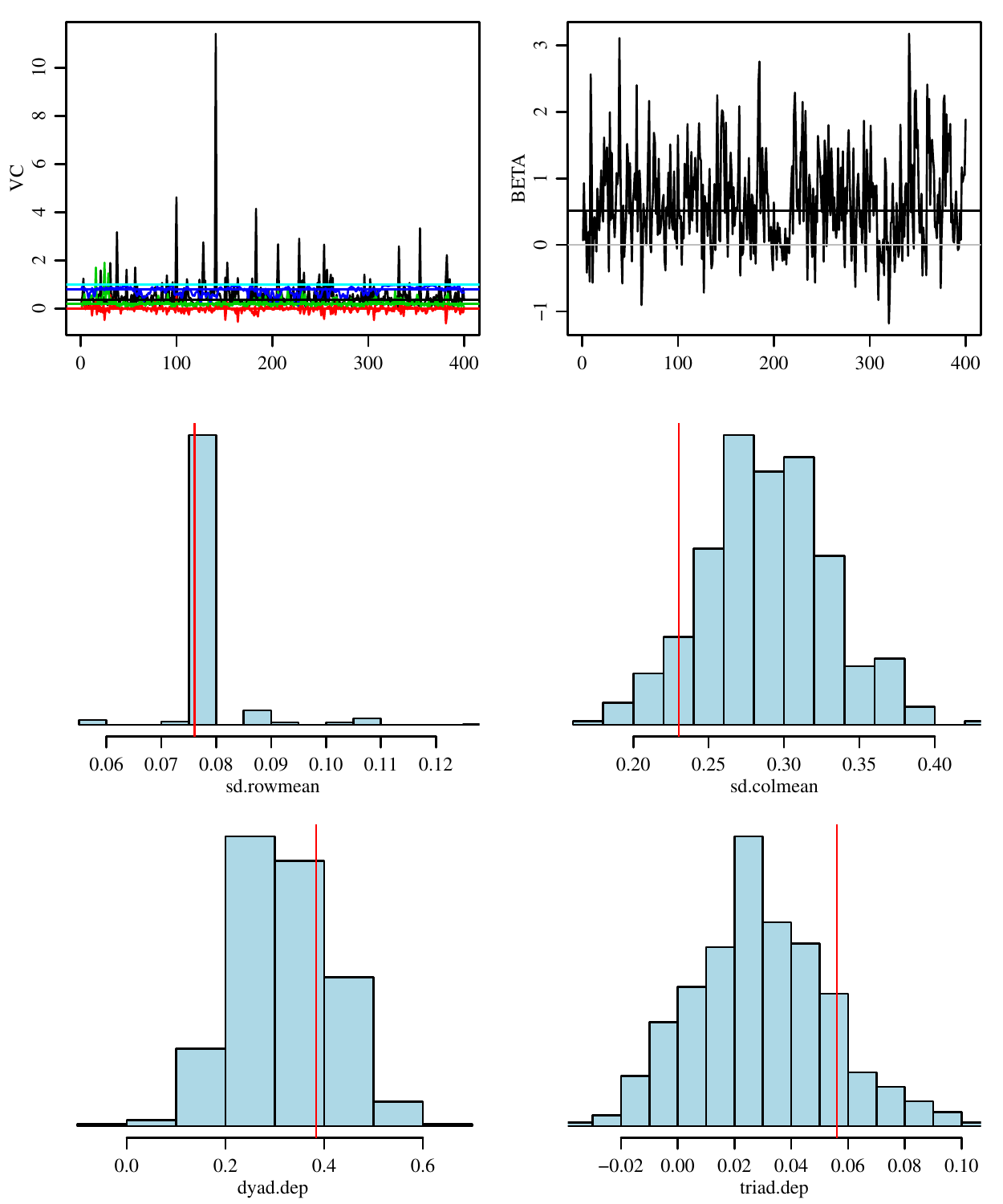} 

}

\end{knitrout}
\caption{Model fitting plots for Sampson's monk data.} 
\label{fig:smonk}
\end{figure}

Goodness of fit plots for these data appear in Figure 
\ref{fig:smonk}. Notice that the fit in terms of row heterogeneity is 
very good. This is not too surprising: 
The simulated sociomatrices used to produce this plot 
are generated to satisfy the constraint imposed 
the outdegree constraint imposed by {\tt odmax}, which 
greatly limits the possible amount of outdegree heterogeneity. 

\subsection{Other approaches to censored or ranked data}
Some dyadic survey designs ask participants to nominate 
up to a certain number of friends, but not to rank them. Such dyadic data
are binary, but censored in the same way as are data from an FRN survey: 
Observing that $y_{i,j}=0$ indicates that $i$ is not friends with $j$ 
only if person $i$ has made less than the maximum number  of nominations. 
A likelihood-based approach to analyzing such 
censored binary data is described in 
\citet{hoff_fosdick_volfovsky_stovel_2013} and 
is also implemented in the {\tt amen} package
using the {\tt model="cbin"} option in the 
{\tt ame} command. 

In other situations the dyadic outcomes in each 
row are ordinal, but on completely different scales. 
In such cases, we may wish to treat the heterogeneity of 
ties across rows in a semiparametric way, and only 
estimate the parameters in the AME model based on the 
ranks of the outcomes within each row.  
This can be done by using a likelihood 
for which the ordinal dyadic data $\bl Y$ only imposes 
constraint (\ref{eqn:rnkc}) on the unobserved underlying 
variables  $\bl Z$.  
This ``relative rank likelihood'' is described more fully in 
\citet{hoff_fosdick_volfovsky_stovel_2013}, 
and can be implemented in {\tt amen} using the {\tt model="rrl"} option in the
{\tt ame} command.

\section{Sampled or missing dyadic data}
Some dyadic datasets are only partially observed, in that 
the value of $y_{i,j}$ is not observed for all pairs 
$i,j$.  This can happen unintentionally or by design. 
For example, to avoid the cost of measuring $y_{i,j}$ for 
all $n(n-1)$ ordered pairs of nodes, some researchers use 
multi-stage link-tracing designs, in which 
nodes are selected into the study in one stage of the 
design based on their links to nodes included in previous 
stages.

Partially observed dyadic data on a given nodeset
can be represented by a sociomatrix in which the 
ordered pairs for which data are not observed  are distinguished 
from pairs for which data are observed. 
In {\sf R}, this is done by 
filling each entry of the sociomatrix 
corresponding to a missing value with an ``{\tt NA}''. 
Doing so distinguishes
pairs $i,j$ for which we do not know the  dyadic value 
$(y_{i,j} = {\tt NA})$ from those, for example,  
for which we know there is no link $(y_{i,j}=0)$. 

When some (non-diagonal) entries of the sociomatrix $\bl Y$ are missing, 
the MCMC approximation algorithm used by {\tt amen}
proceeds by iteratively simulating
model parameters 
along with 
values for the missing values in a way that 
approximates their joint posterior distribution.
Roughly speaking, at each iteration  of the MCMC algorithm, 
values for the missing values are simulated from 
their probability distribution conditional on the observed 
data and the current values of the model parameters. 
Such a procedure is appropriate if the missing values 
are \emph{missing at random}, or more specifically, if the 
study design is \emph{ignorable}. 
A study design is ignorable if 
the probability of a missing data value for a pair is 
independent of the model parameters and missing 
values, conditional on the observed data values. 
Many types of link tracing designs, such as egocentric and snowball sampling, 
are ignorable \citep{thompson_frank_2000}. 

\subsection{Example: Analysis of an egocentric sample}
One popular and relatively inexpensive design for gathering 
dyadic data is with an egocentric sample, in which nodes 
(or ``egos'') are randomly sampled from a population and then 
asked about their ties and the ties between their friends. For example, 
one type of egocentric study design might ask participants
``with whom are you friends'' and ``which of your friends are friends 
with each other.''  Data from such a design can be sufficient to estimate 
parameters in an AME model. 

We illustrate this with an example analysis of the effect of 
sex (male/female) on friendships among a small group of Dutch college 
students, available in {\tt amen} from the {\tt dutchcollege} dataset. 
\begin{knitrout}\footnotesize
\definecolor{shadecolor}{rgb}{0.969, 0.969, 0.969}\color{fgcolor}\begin{kframe}
\begin{alltt}
\hlkwd{data}\hlstd{(dutchcollege)}

\hlstd{Y}\hlkwb{<-}\hlnum{1}\hlopt{*}\hlstd{( dutchcollege}\hlopt{$}\hlstd{Y[,,}\hlnum{7}\hlstd{]} \hlopt{>} \hlnum{1} \hlstd{)} \hlcom{# indicator of positive relationship at the last timepoint}

\hlstd{Xn}\hlkwb{<-}\hlstd{dutchcollege}\hlopt{$}\hlstd{X[,}\hlnum{1}\hlstd{]}           \hlcom{# nodal indicator of male sex}
\hlstd{Xd}\hlkwb{<-}\hlnum{1}\hlopt{*}\hlstd{(}\hlkwd{outer}\hlstd{(Xn,Xn,}\hlstr{"=="}\hlstd{))}        \hlcom{# dyadic indicator of same sex}
\end{alltt}
\end{kframe}
\end{knitrout}

We will fit a simple SRRM to these data, and then compare the resulting parameter
estimates to those based on data obtained 
from the egocentric design described above.
In our design, we 
first randomly sample several nodes (egos), 
record their relationships to the other nodes, and then 
record the relationships between alters having a common ego. 
{\sf R}-code that generates such a design is as follows:

\begin{knitrout}\footnotesize
\definecolor{shadecolor}{rgb}{0.969, 0.969, 0.969}\color{fgcolor}\begin{kframe}
\begin{alltt}
\hlstd{n}\hlkwb{<-}\hlkwd{nrow}\hlstd{(Y)}
\hlstd{Ys}\hlkwb{<-}\hlkwd{matrix}\hlstd{(}\hlnum{NA}\hlstd{,n,n)}      \hlcom{# sociomatrix for sampled data}

\hlstd{egos}\hlkwb{<-}\hlkwd{sort}\hlstd{(}\hlkwd{sample}\hlstd{(n,}\hlnum{5}\hlstd{))} \hlcom{# ego sample}
\hlstd{Ys[egos,]}\hlkwb{<-}\hlstd{Y[egos,]}     \hlcom{# relations of egos are observed}

\hlkwa{for}\hlstd{(i} \hlkwa{in} \hlstd{egos)}
\hlstd{\{}
  \hlstd{ai}\hlkwb{<-}\hlkwd{which}\hlstd{(Ys[i,]}\hlopt{==}\hlnum{1}\hlstd{)}  \hlcom{# alters of i  }
  \hlstd{Ys[ai,ai]}\hlkwb{<-}\hlstd{Y[ai,ai]}   \hlcom{# relations between alters of i are observed}
\hlstd{\}}

\hlkwd{mean}\hlstd{(}\hlkwd{is.na}\hlstd{(Ys))}
\end{alltt}
\begin{verbatim}
[1] 0.7314453
\end{verbatim}
\end{kframe}
\end{knitrout}
This particular instance of the design results in a sociomatrix 
where about 73 
percent of the entries are missing
(note that the diagonal is  already ``missing'' by definition). 
Under this design, data between alters and non-alters of an ego
are missing, as are data between alters that do not share an ego.
\begin{knitrout}\footnotesize
\definecolor{shadecolor}{rgb}{0.969, 0.969, 0.969}\color{fgcolor}\begin{kframe}
\begin{alltt}
\hlstd{egos}
\end{alltt}
\begin{verbatim}
[1]  6 10 12 17 26
\end{verbatim}
\begin{alltt}
\hlstd{Ys[}\hlnum{1}\hlopt{:}\hlnum{10}\hlstd{,}\hlnum{1}\hlopt{:}\hlnum{10}\hlstd{]}
\end{alltt}
\begin{verbatim}
      [,1] [,2] [,3] [,4] [,5] [,6] [,7] [,8] [,9] [,10]
 [1,]   NA   NA   NA   NA   NA   NA   NA   NA   NA    NA
 [2,]   NA   NA    0   NA   NA   NA    0    0   NA     1
 [3,]   NA    0   NA   NA   NA   NA    1    0   NA    NA
 [4,]   NA   NA   NA   NA   NA   NA   NA   NA   NA    NA
 [5,]   NA   NA   NA   NA   NA   NA   NA   NA   NA    NA
 [6,]    0    0    0    0    0   NA    0    0    0     0
 [7,]   NA    0    1   NA   NA   NA   NA    1   NA    NA
 [8,]   NA    0    0   NA   NA   NA    0   NA   NA     1
 [9,]   NA   NA   NA   NA   NA   NA   NA   NA   NA    NA
[10,]    0    1    1    0    0    0    1    1    0    NA
\end{verbatim}
\end{kframe}
\end{knitrout}

We now fit an SRRM  model to the complete data and the subsampled data, and compare 
parameter estimates. 
\begin{knitrout}\footnotesize
\definecolor{shadecolor}{rgb}{0.969, 0.969, 0.969}\color{fgcolor}\begin{kframe}
\begin{alltt}
\hlstd{fit_pop}\hlkwb{<-}\hlkwd{ame}\hlstd{(Y,Xd,Xn,Xn,}\hlkwc{model}\hlstd{=}\hlstr{"bin"}\hlstd{)}   \hlcom{# fit based on full data (population) }

\hlstd{fit_ess}\hlkwb{<-}\hlkwd{ame}\hlstd{(Ys,Xd,Xn,Xn,}\hlkwc{model}\hlstd{=}\hlstr{"bin"}\hlstd{)}  \hlcom{# fit based on egocentric subsample}
\end{alltt}
\end{kframe}
\end{knitrout}

\begin{knitrout}\footnotesize
\definecolor{shadecolor}{rgb}{0.969, 0.969, 0.969}\color{fgcolor}\begin{kframe}
\begin{alltt}
\hlkwd{apply}\hlstd{(fit_pop}\hlopt{$}\hlstd{BETA,}\hlnum{2}\hlstd{,mean)}
\end{alltt}
\begin{verbatim}
 intercept       .row       .col      .dyad 
-1.8662999  0.3172635  0.3824924  0.7365514 
\end{verbatim}
\begin{alltt}
\hlkwd{apply}\hlstd{(fit_ess}\hlopt{$}\hlstd{BETA,}\hlnum{2}\hlstd{,mean)}
\end{alltt}
\begin{verbatim}
 intercept       .row       .col      .dyad 
-2.1561520  0.3172207  1.1966069  0.9929780 
\end{verbatim}
\end{kframe}
\end{knitrout}
\noindent
The estimates  are similar, even though the second fit is from  
a dataset with 73 
missing values. 


The output of the {\tt ame} fitting procedure also includes a posterior 
predictive mean for all entries of the sociomatrix $\bl Y$, 
including those entries for which the data are missing. 
This sociomatrix of predicted values can be used for prediction 
or imputation of dyadic data from 
incomplete datasets. In our example on modeling friendship 
relations from the {\tt dutchcollege} dataset, we can 
use this sociomatrix of predicted values to evaluate 
how well the parameter estimates obtained from  the sampled dataset  
compare to those obtained from the full dataset, in terms of 
prediction:
\begin{knitrout}\footnotesize
\definecolor{shadecolor}{rgb}{0.969, 0.969, 0.969}\color{fgcolor}\begin{kframe}
\begin{alltt}
\hlstd{miss}\hlkwb{<-}\hlkwd{which}\hlstd{(}\hlkwd{is.na}\hlstd{(Ys))}

\hlkwd{mean}\hlstd{( ( fit_pop}\hlopt{$}\hlstd{YPM[miss]} \hlopt{-} \hlstd{Y[miss] )}\hlopt{^}\hlnum{2}\hlstd{,} \hlkwc{na.rm}\hlstd{=}\hlnum{TRUE} \hlstd{)}
\end{alltt}
\begin{verbatim}
[1] 0.09428416
\end{verbatim}
\begin{alltt}
\hlkwd{mean}\hlstd{( ( fit_ess}\hlopt{$}\hlstd{YPM[miss]} \hlopt{-} \hlstd{Y[miss] )}\hlopt{^}\hlnum{2}\hlstd{,} \hlkwc{na.rm}\hlstd{=}\hlnum{TRUE} \hlstd{)}
\end{alltt}
\begin{verbatim}
[1] 0.1350285
\end{verbatim}
\end{kframe}
\end{knitrout}
The first and second numbers reflect ``within-sample'' 
and ``out-of-sample'' goodness of fit, respectively. The 
small discrepancy between these numbers indicates that
reasonable parameter estimates  for this model 
(in terms of out-of-sample predictive squared error) can be obtained from 
this egocentric sample.

\begin{figure}
\begin{knitrout}\footnotesize
\definecolor{shadecolor}{rgb}{0.969, 0.969, 0.969}\color{fgcolor}

{\centering \includegraphics[width=\maxwidth]{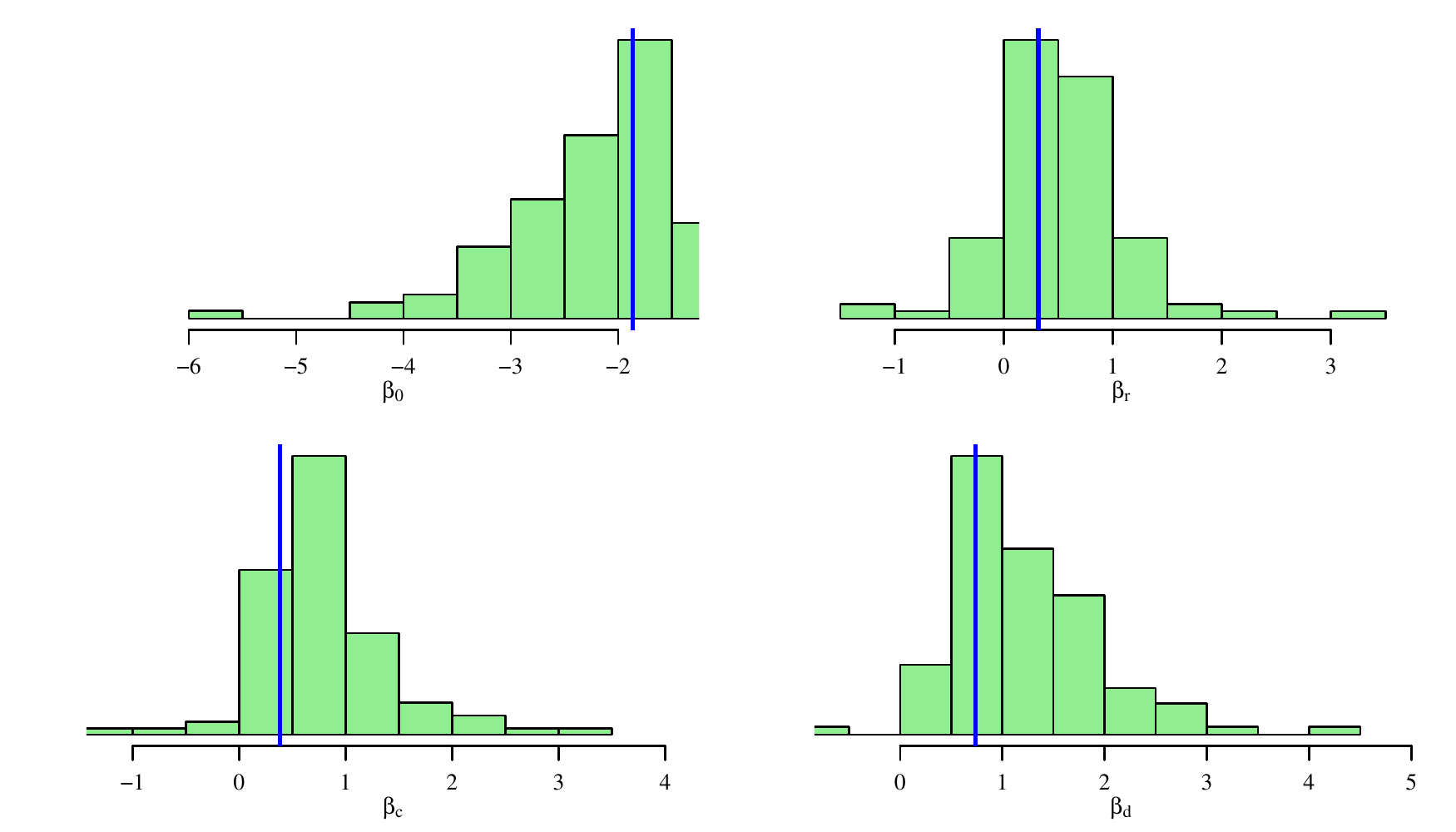} 

}

\end{knitrout}
\caption{Variability of probit SRRM regression estimates across 
egocentric samples. Vertical blue  lines indicate the estimates 
obtained from the full (population) dataset. }
\label{fig:ess_betas}
\end{figure}

Finally, we consider the variability of the  parameter estimates 
across egocentric samples with a small simulation study: For each 
of 100 egocentric samples randomly generated as previously described, we obtain 
parameter estimates for the probit SRRM, 
\begin{align*} 
z_{i,j} &= \beta_0 +\beta_r x_i+\beta_c x_j +\beta_d x_{i,j}+\epsilon_{i,j}\\
y_{i,j} &= 1( z_{i,j}>0 ),
\end{align*}
where $x_i$ is a binary indicator that node $i$ is male, 
and $x_{i,j}$ is the indicator that $i$ and $j$ are of the same sex. 
The variability of the parameter estimates across
egocentric samples is illustrated with histograms in 
Figure \ref{fig:ess_betas}. An illustrative exercise would be to 
see how increasing or decreasing the amount of missing data in the samples 
(by increasing or decreasing the number of egos sampled) would 
affect the concentration of the egocentric estimates around the 
population estimates. 

\section{Repeated measures data}
Some types of dynamic dyadic datasets include 
repeated measurements of dyadic 
and nodal variables at discrete points in time. 
The {\tt amen} package provides a rudimentary 
method of analyzing such data, 
based on the following simple extension of the 
AME model to accommodate 
replicated dyadic measurements:
For (latent) sociomatrices $\bl Z_1,\ldots, \bl Z_T$, 
the model expresses $z_{i,j,t}$, the $(i,j)$th element of the
$t$th sociomatrix, as 
\begin{equation} 
z_{i,j,t} =  \beta_d^T\bl x_{d,i,j,t} +
 \beta_r^T\bl x_{r,i,j,t} +\beta_c^T\bl x_{c,i,j,t} + 
a_i +b_j + \bl u_i^T \bl v_j + \epsilon_{i,j,t}. 
\label{eqn:llame}
\end{equation}
Across nodes, dyads and time points,  this model extension  further
assumes the same covariance model for the random effects  $\{(a_i,b_i)\}$ 
as before, 
allows for 
dyadic correlation between $\epsilon_{i,j,t}$ and  
$\epsilon_{j,i,t}$ as before, but 
assumes that
$\epsilon_{i,j,t}$'s from 
different dyads \emph{or} different time points are independent. 
In other words, the data under this 
model are treated as 
independent observations from a common AME distribution. 

At first glance it may seem that such a model is inappropriate for 
dynamic dyadic data, as it doesn't seem to allow for the 
possibility of dependence over time. 
However, 
certain types of dependence can be incorporated into this model 
via the 
time-dependent 
regression terms. For example, autoregressive
dependence can be modeled by including lagged values of the 
sociomatrix as predictors. Additionally, 
time-varying regression parameters can be included in the 
model by constructing interactions.

\subsection{Example: Analysis of a longitudinal binary outcome}

\begin{figure}
\begin{knitrout}\footnotesize
\definecolor{shadecolor}{rgb}{0.969, 0.969, 0.969}\color{fgcolor}
\includegraphics[width=\maxwidth]{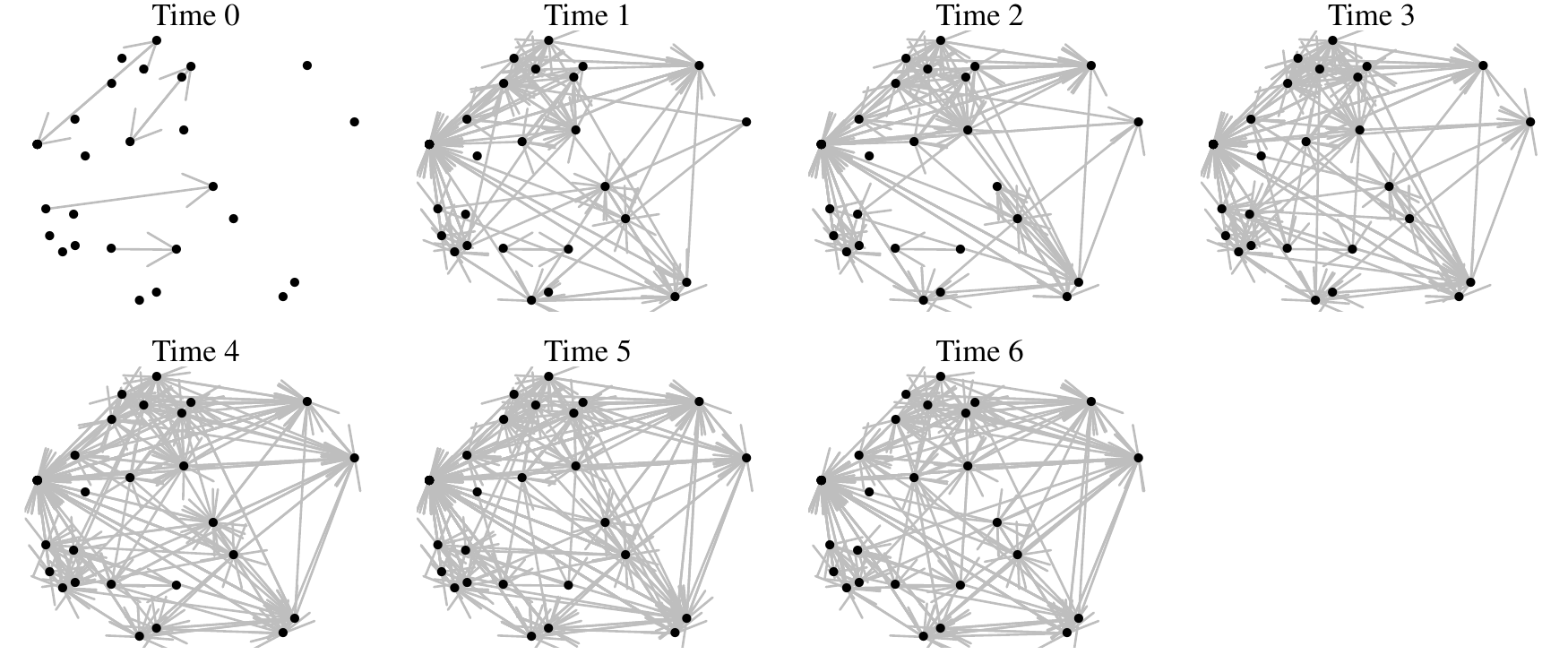} 

\end{knitrout}
\caption{Friendliness network of the Dutch college students, 
across seven time points.} 
\label{fig:dcln}
\end{figure}

We illustrate these possibilities with an analysis 
of data from the longitudinal study of friendship relations 
among a small group of Dutch college students, available 
in {\tt amen} via the {\tt dutchcollege} dataset. 
Our response $y_{i,j,t}$ is the indicator that  
person $i$ reports being friendly (or having friendship) 
with person $j$ at time point 
$t$. 
The graphs of this variable for each of  the seven different time points 
in the dataset are given in 
Figure \ref{fig:dcln}. 
The figure reflects the fact that 
the students were mostly unknown to each other before the 
study period, and so not surprisingly, the densities of the graphs
increase over time. 

The data also include (static) information on 
the sex and smoking status of the students, as 
well as which one of three programs each student was a member. 
We will examine the effects of these nodal 
attributes on friendship in using a probit 
SRRM, where $y_{i,j,t}$ is modeled 
as the indicator that the latent affinity 
$z_{i,j,t}$ exceeds zero, where 
$z_{i,j,t}$ follows model 
(\ref{eqn:llame}). 
Our analysis will include the 
binary indicators of male sex and 
smoking status as row and column regressors, 
products of these variables as dyadic regressors, 
and a dyadic binary indicator of whether or 
not members of a dyad belong to the same program.  
Finally, we will also include lagged values 
$y_{i,j,t-1}$ and $y_{j,i,t-1}$ as dyadic predictors
of $z_{i,j,t}$ to
reflect the possibility of temporal dependence among 
values  within a dyad.  To summarize, our model 
for the $z_{i,j,t}$'s is as follows:
\begin{align*}
 z_{i,j,t}  = &   \beta_0 +  \\ 
 & \beta_{r,1} \text{male}_i + \beta_{r,2} \text{smoke}_i +  \\
 & \beta_{c,1} \text{male}_j + \beta_{c,2} \text{smoke}_j +  \\
 & \beta_{d_1} y_{i,j,t-1} + \beta_{d_2} y_{j,i,t-1} +  \\
 & \beta_{d_3} \text{male}_i \text{male}_j +
  \beta_{d_4} \text{smoke}_i \text{smoke}_j + 
  \beta_{d_5} \text{sameprogram}_{i,j} +  \\
 &  a_i +b_j + \epsilon_{i,j,t} 
\end{align*}

The {\tt ame\_rep} function in the {\tt amen} package 
provides parameter estimation and inference  for this model
using a similar syntax as the {\tt ame} function, 
except now the 
nodal attributes {\tt Xrow} and {\tt Xcol} are
three-dimensional arrays, with dimensions corresponding 
to nodes, variables and time points, respectively. 
Similarly, 
the dyadic regressor array {\tt Xdyad} is now four-dimensional, 
 with dimensions corresponding 
to nodes, nodes,  variables and time points. 
These arrays for this data analysis can be set up as follows:
\begin{knitrout}\footnotesize
\definecolor{shadecolor}{rgb}{0.969, 0.969, 0.969}\color{fgcolor}\begin{kframe}
\begin{alltt}
\hlcom{# outcome}
\hlstd{Y}\hlkwb{<-}\hlnum{1}\hlopt{*}\hlstd{( dutchcollege}\hlopt{$}\hlstd{Y} \hlopt{>=} \hlnum{2} \hlstd{)[,,}\hlnum{2}\hlopt{:}\hlnum{7}\hlstd{]}
\hlstd{n}\hlkwb{<-}\hlkwd{dim}\hlstd{(Y)[}\hlnum{1}\hlstd{] ; t}\hlkwb{<-}\hlkwd{dim}\hlstd{(Y)[}\hlnum{3}\hlstd{]}

\hlcom{# nodal covariates}
\hlstd{Xnode}\hlkwb{<-}\hlstd{dutchcollege}\hlopt{$}\hlstd{X[,}\hlnum{1}\hlopt{:}\hlnum{2}\hlstd{]}                                       \hlcom{# sex and smoking status}
\hlstd{Xnode}\hlkwb{<-}\hlkwd{array}\hlstd{(Xnode,}\hlkwc{dim}\hlstd{=}\hlkwd{c}\hlstd{(n,}\hlkwd{ncol}\hlstd{(Xnode),t))}
\hlkwd{dimnames}\hlstd{(Xnode)[[}\hlnum{2}\hlstd{]]}\hlkwb{<-}\hlkwd{c}\hlstd{(}\hlstr{"male"}\hlstd{,}\hlstr{"smoker"}\hlstd{)}

\hlcom{# dyadic covariates}
\hlstd{Xdyad}\hlkwb{<-}\hlkwd{array}\hlstd{(}\hlkwc{dim}\hlstd{=}\hlkwd{c}\hlstd{(n,n,}\hlnum{5}\hlstd{,t))}
\hlstd{Xdyad[,,}\hlnum{1}\hlstd{,]}\hlkwb{<-}\hlnum{1}\hlopt{*}\hlstd{( dutchcollege}\hlopt{$}\hlstd{Y} \hlopt{>=} \hlnum{2} \hlstd{)[,,}\hlnum{1}\hlopt{:}\hlnum{6}\hlstd{]}                     \hlcom{# lagged value}
\hlstd{Xdyad[,,}\hlnum{2}\hlstd{,]}\hlkwb{<-}\hlkwd{array}\hlstd{(}\hlkwd{apply}\hlstd{(Xdyad[,,}\hlnum{1}\hlstd{,],}\hlnum{3}\hlstd{,t),}\hlkwc{dim}\hlstd{=}\hlkwd{c}\hlstd{(n,n,t))}           \hlcom{# lagged reciprocal value }
\hlstd{Xdyad[,,}\hlnum{3}\hlstd{,]}\hlkwb{<-}\hlkwd{tcrossprod}\hlstd{(Xnode[,}\hlnum{1}\hlstd{,}\hlnum{1}\hlstd{])}                              \hlcom{# both male}
\hlstd{Xdyad[,,}\hlnum{4}\hlstd{,]}\hlkwb{<-}\hlkwd{tcrossprod}\hlstd{(Xnode[,}\hlnum{2}\hlstd{,}\hlnum{1}\hlstd{])}                              \hlcom{# both smokers }
\hlstd{Xdyad[,,}\hlnum{5}\hlstd{,]}\hlkwb{<-}\hlkwd{outer}\hlstd{( dutchcollege}\hlopt{$}\hlstd{X[,}\hlnum{3}\hlstd{],dutchcollege}\hlopt{$}\hlstd{X[,}\hlnum{3}\hlstd{],}\hlstr{"=="}\hlstd{)}   \hlcom{# same program}
\hlkwd{dimnames}\hlstd{(Xdyad)[[}\hlnum{3}\hlstd{]]}\hlkwb{<-}\hlkwd{c}\hlstd{(}\hlstr{"Ylag"}\hlstd{,}\hlstr{"tYlag"}\hlstd{,}\hlstr{"bothmale"}\hlstd{,}\hlstr{"bothsmoke"}\hlstd{,}\hlstr{"sameprog"}\hlstd{)}
\end{alltt}
\end{kframe}
\end{knitrout}

The model can be fit using the same syntax as the 
{\tt ame} command discussed previously, and results can 
summarized before with the {\tt summary} function:
\begin{knitrout}\footnotesize
\definecolor{shadecolor}{rgb}{0.969, 0.969, 0.969}\color{fgcolor}\begin{kframe}
\begin{alltt}
\hlstd{fit_ar1}\hlkwb{<-}\hlkwd{ame_rep}\hlstd{(Y,Xdyad,Xnode,Xnode,}\hlkwc{model}\hlstd{=}\hlstr{"bin"}\hlstd{,}\hlkwc{plot}\hlstd{=}\hlnum{FALSE}\hlstd{)}
\end{alltt}
\end{kframe}
\end{knitrout}

\begin{knitrout}\footnotesize
\definecolor{shadecolor}{rgb}{0.969, 0.969, 0.969}\color{fgcolor}\begin{kframe}
\begin{alltt}
\hlkwd{summary}\hlstd{(fit_ar1)}
\end{alltt}
\begin{verbatim}

Regression coefficients:
                pmean   psd z-stat p-val
intercept      -1.612 0.170 -9.457 0.000
male.row       -0.170 0.220 -0.772 0.440
smoker.row     -0.458 0.182 -2.516 0.012
male.col       -0.038 0.162 -0.236 0.813
smoker.col     -0.236 0.145 -1.627 0.104
Ylag.dyad       1.201 0.063 19.146 0.000
tYlag.dyad      0.860 0.062 13.796 0.000
bothmale.dyad   0.740 0.145  5.090 0.000
bothsmoke.dyad  0.661 0.122  5.424 0.000
sameprog.dyad   0.432 0.063  6.880 0.000

Variance parameters:
    pmean   psd
va  0.223 0.073
cab 0.033 0.034
vb  0.119 0.038
rho 0.641 0.038
ve  1.000 0.000
\end{verbatim}
\end{kframe}
\end{knitrout}
The 
parameter estimates and standard deviations for 
$\beta_{d,1}$ and $\beta_{d,2}$ ({\tt Ylag.dyad} and 
{\tt tYlag.dyad} in the output) 
indicate strong evidence 
of large 
temporal correlation. 
There also appears to be strong homophily effects in 
terms of sex, smoking status and program. 
The nodal effect parameters indicate some evidence 
that smokers are a bit less social than non-smokers. 

Finally, we note that the time interval between 
the first four measurements was three weeks, whereas the 
interval between the last three measurements was six weeks. 
As such, we may want to consider whether or not the 
effects of the regressors might vary depending on the 
time lag between measurements. Such a possibility can be evaluated simply 
by adding interaction terms to the regressors. 
For example, to evaluate whether or not the effect of 
$y_{i,j,t-1}$ on $z_{i,j,t}$ varies with measurement interval, 
we can create a new dyadic covariate 
$y_{i,j,t-1} w_{t}$ where 
$w_t$ is a binary indicator that $t$ is 
among the last three measurements. 
Adding such terms for all of our regressors can be done as follows:
\begin{knitrout}\footnotesize
\definecolor{shadecolor}{rgb}{0.969, 0.969, 0.969}\color{fgcolor}\begin{kframe}
\begin{alltt}
\hlstd{Wnode}\hlkwb{<-}\hlstd{Xnode}
\hlstd{Wnode[,,}\hlnum{1}\hlopt{:}\hlnum{3}\hlstd{]}\hlkwb{<-}\hlnum{0}

\hlstd{XWnode}\hlkwb{<-}\hlkwd{array}\hlstd{(} \hlkwc{dim}\hlstd{=}\hlkwd{dim}\hlstd{(Xnode)}\hlopt{+}\hlkwd{c}\hlstd{(}\hlnum{0}\hlstd{,}\hlnum{2}\hlstd{,}\hlnum{0}\hlstd{))}
\hlstd{XWnode[,}\hlnum{1}\hlopt{:}\hlnum{2}\hlstd{,]}\hlkwb{<-}\hlstd{Xnode ; XWnode[,}\hlnum{3}\hlopt{:}\hlnum{4}\hlstd{,]}\hlkwb{<-}\hlstd{Wnode}
\hlkwd{dimnames}\hlstd{(XWnode)[[}\hlnum{2}\hlstd{]]}\hlkwb{<-}\hlkwd{c}\hlstd{(}\hlkwd{dimnames}\hlstd{(Xnode)[[}\hlnum{2}\hlstd{]],}\hlkwd{paste0}\hlstd{(}\hlkwd{dimnames}\hlstd{(Xnode)[[}\hlnum{2}\hlstd{]],}\hlstr{".w"}\hlstd{))}


\hlstd{Wdyad}\hlkwb{<-}\hlstd{Xdyad}
\hlstd{Wdyad[,,,}\hlnum{1}\hlopt{:}\hlnum{3}\hlstd{]}\hlkwb{<-}\hlnum{0}
\hlstd{XWdyad}\hlkwb{<-}\hlkwd{array}\hlstd{(} \hlkwc{dim}\hlstd{=}\hlkwd{dim}\hlstd{(Xdyad)}\hlopt{+}\hlkwd{c}\hlstd{(}\hlnum{0}\hlstd{,}\hlnum{0}\hlstd{,}\hlnum{5}\hlstd{,}\hlnum{0}\hlstd{) )}
\hlstd{XWdyad[,,}\hlnum{1}\hlopt{:}\hlnum{5}\hlstd{,]}\hlkwb{<-}\hlstd{Xdyad ; XWdyad[,,}\hlnum{6}\hlopt{:}\hlnum{10}\hlstd{,]}\hlkwb{<-}\hlstd{Wdyad}
\hlkwd{dimnames}\hlstd{(XWdyad)[[}\hlnum{3}\hlstd{]]}\hlkwb{<-}\hlkwd{c}\hlstd{(}\hlkwd{dimnames}\hlstd{(Xdyad)[[}\hlnum{3}\hlstd{]],}\hlkwd{paste0}\hlstd{(}\hlkwd{dimnames}\hlstd{(Xdyad)[[}\hlnum{3}\hlstd{]],}\hlstr{".w"}\hlstd{))}
\end{alltt}
\end{kframe}
\end{knitrout}

\begin{knitrout}\footnotesize
\definecolor{shadecolor}{rgb}{0.969, 0.969, 0.969}\color{fgcolor}\begin{kframe}
\begin{alltt}
\hlstd{fit_ar1_vb}\hlkwb{<-}\hlkwd{ame_rep}\hlstd{(Y,XWdyad,XWnode,XWnode,}\hlkwc{model}\hlstd{=}\hlstr{"bin"}\hlstd{)}
\end{alltt}
\end{kframe}
\end{knitrout}

\begin{knitrout}\footnotesize
\definecolor{shadecolor}{rgb}{0.969, 0.969, 0.969}\color{fgcolor}\begin{kframe}
\begin{alltt}
\hlkwd{summary}\hlstd{(fit_ar1_vb)}
\end{alltt}
\begin{verbatim}

Regression coefficients:
                  pmean   psd z-stat p-val
intercept        -1.606 0.167 -9.625 0.000
male.row         -0.313 0.238 -1.311 0.190
smoker.row       -0.374 0.204 -1.833 0.067
male.w.row        0.240 0.141  1.696 0.090
smoker.w.row     -0.156 0.137 -1.134 0.257
male.col         -0.068 0.171 -0.395 0.693
smoker.col       -0.208 0.148 -1.402 0.161
male.w.col        0.036 0.133  0.268 0.788
smoker.w.col     -0.048 0.120 -0.403 0.687
Ylag.dyad         1.400 0.101 13.797 0.000
tYlag.dyad        0.855 0.109  7.839 0.000
bothmale.dyad     1.009 0.209  4.831 0.000
bothsmoke.dyad    0.558 0.165  3.373 0.001
sameprog.dyad     0.334 0.080  4.169 0.000
Ylag.w.dyad      -0.296 0.122 -2.432 0.015
tYlag.w.dyad     -0.008 0.131 -0.059 0.953
bothmale.w.dyad  -0.520 0.277 -1.875 0.061
bothsmoke.w.dyad  0.194 0.225  0.864 0.387
sameprog.w.dyad   0.187 0.097  1.924 0.054

Variance parameters:
    pmean   psd
va  0.224 0.075
cab 0.032 0.036
vb  0.116 0.038
rho 0.650 0.037
ve  1.000 0.000
\end{verbatim}
\end{kframe}
\end{knitrout}
These results do not indicate much evidence that the 
regression coefficients should vary by time period, 
except possibly the effect on the lagged 
dyadic variable $y_{i,j,t}$. The negative 
estimate of this coefficient 
(corresponding to {\tt Ylag.w.dyad} in the output) 
makes sense, 
as it 
indicates  that the effect of the lagged variable is 
decreased when the interval between times points is longer.

\section{Symmetric data}
It is sometimes the case that dyadic
observations are \emph{symmetric} or 
\emph{undirected} by design, in that there is 
only one  value $y_{i,j}$ for the 
dyad $\{i,j\}$. Such observations can 
be represented by a 
symmetric sociomatrix $\bl Y$, so that 
$y_{i,j}= y_{j,i}$ for all dyads $\{i,j\}$. 
In this case, 
a natural simplification of the AME  model 
(\ref{eqn:ame})  is given by 
\begin{align}
 y_{i,j}  & = \beta_d^T \bl x_{i,j} + \beta_n^T ( \bl x_{i} + \bl x_{j} ) +  a_i + a_j +  \bl u_i^T \bl \Lambda  \bl u_j +  \epsilon_{i,j} ,  \label{eqn:same}  \\ 
 a_1,\ldots, a_n & \sim  \text{i.i.d.} \  N(0,\sigma^2_a) \nonumber  \\
 \{ \epsilon_{i,j} \}  & \sim  \text{i.i.d.}  \ N(0,\sigma^2_e),  \nonumber
\end{align}
for $i<j$, with $y_{j,i} = y_{i,j}$ by design. 
Most of the simplifications leading to this symmetric model are 
easy to understand, with the possible exception of the 
change from $\bl u_i^T\bl v_j$ in the asymmetric case 
to $\bl u_i^T \bs \Lambda \bl u_j$ here. 
In the former case, this representation can be justified 
by the singular value decomposition theorem, 
which states that
any $n\times n$ rank-$R$ matrix $\bl M$ can be expressed 
as $\bl U \bl V^T$, where 
$\bl U$ and $\bl V$ are  ${n\times R}$ matrices. 
This means that $m_{i,j}$, the $i,j$th entry of $\bl M$ 
can be expressed as $m_{i,j} = \bl u_i^T \bl v_j$, where 
$\bl u_i$ and $\bl v_j$ are the $i$th and $j$th rows of 
$\bl U$ and $\bl V$, respectively. 
In other words, the $\bl u_i^T\bl v_j$ 
in the asymmetric AME model can represent any residual 
low-rank patterns $\bl M$ in the sociomatrix  $\bl Y$ that aren't explained 
by the known regressors. 
Similarly, in the symmetric case 
the term $\bl u_i^T \bs \Lambda \bl u_j$ in 
(\ref{eqn:same}) can represent any residual low-rank patterns 
$\bl M$  in the symmetric sociomatrix $\bl Y$. 
This follows from the eigenvalue decomposition theorem, 
which states that 
any symmetric rank-$R$
matrix $\bl M$ can be expressed as $\bl U \bs \Lambda \bl U^T$, or 
equivalently, 
the elements $m_{i,j}$ of $\bl M$ 
can be expressed as 
$m_{i,j} = \bl  u_i^T \bs \Lambda \bl u_j$. 
Furthermore, as with the asymmetric case, 
such a latent factor model can
represent patterns of transitivity and stochastic equivalence 
in network data \citep{hoff_2008}. 

\subsection{Example: Analysis of a symmetric ordinal outcome}

Symmetric versions of the  normal, probit and other AME models discussed  
in the previous sections can be fit 
by simply specifying the option  {\tt symmetric=TRUE} in the 
 {\tt ame} command. 
We illustrate the use of this option with an analysis 
of Cold War cooperation and conflict data, available 
via the {\tt coldwar} dataset. 
These data include 
dyadic 
 counts of military cooperation and conflict between countries, 
geographic distances between countries, 
and country-level measures of GDP and polity. 
These variables were recorded every five years from 1950 to 1985. 
For simplicity, we analyze a time-averaged version of the dataset:
\begin{knitrout}\footnotesize
\definecolor{shadecolor}{rgb}{0.969, 0.969, 0.969}\color{fgcolor}\begin{kframe}
\begin{alltt}
\hlkwd{data}\hlstd{(coldwar)}

\hlcom{# response}
\hlstd{Y}\hlkwb{<-}\hlkwd{sign}\hlstd{(} \hlkwd{apply}\hlstd{(coldwar}\hlopt{$}\hlstd{cc,}\hlkwd{c}\hlstd{(}\hlnum{1}\hlstd{,}\hlnum{2}\hlstd{), mean ) )}

\hlcom{# nodal covariates}
\hlstd{Xn}\hlkwb{<-}\hlkwd{cbind}\hlstd{(} \hlkwd{apply}\hlstd{(} \hlkwd{log}\hlstd{(coldwar}\hlopt{$}\hlstd{gdp),}\hlnum{1}\hlstd{,mean ) ,}       \hlcom{# log gdp}
           \hlkwd{sign}\hlstd{(}\hlkwd{apply}\hlstd{(coldwar}\hlopt{$}\hlstd{polity ,}\hlnum{1}\hlstd{,mean ) ) )}  \hlcom{# sign of polity      }
\hlstd{Xn[,}\hlnum{1}\hlstd{]}\hlkwb{<-}\hlstd{Xn[,}\hlnum{1}\hlstd{]}\hlopt{-}\hlkwd{mean}\hlstd{(Xn[,}\hlnum{1}\hlstd{])}
\hlkwd{dimnames}\hlstd{(Xn)[[}\hlnum{2}\hlstd{]]}\hlkwb{<-}\hlkwd{c}\hlstd{(}\hlstr{"lgdp"}\hlstd{,}\hlstr{"polity"}\hlstd{)}

\hlcom{# dyadic covariates}
\hlstd{Xd}\hlkwb{<-}\hlkwd{array}\hlstd{(}\hlkwc{dim}\hlstd{=}\hlkwd{c}\hlstd{(}\hlkwd{nrow}\hlstd{(Y),}\hlkwd{nrow}\hlstd{(Y),}\hlnum{3}\hlstd{))}
\hlstd{Xd[,,}\hlnum{1}\hlstd{]}\hlkwb{<-} \hlkwd{tcrossprod}\hlstd{(Xn[,}\hlnum{1}\hlstd{])}                        \hlcom{# gdp interaction}
\hlstd{Xd[,,}\hlnum{2}\hlstd{]}\hlkwb{<-} \hlkwd{tcrossprod}\hlstd{(Xn[,}\hlnum{2}\hlstd{])}                        \hlcom{# polity interaction}
\hlstd{Xd[,,}\hlnum{3}\hlstd{]}\hlkwb{<-}\hlkwd{log}\hlstd{(coldwar}\hlopt{$}\hlstd{distance)}                      \hlcom{# log distance}
\hlkwd{dimnames}\hlstd{(Xd)[[}\hlnum{3}\hlstd{]]}\hlkwb{<-}\hlkwd{c}\hlstd{(}\hlstr{"igdp"}\hlstd{,}\hlstr{"ipol"}\hlstd{,}\hlstr{"ldist"}\hlstd{)}
\end{alltt}
\end{kframe}
\end{knitrout}

The response $y_{i,j}$ takes values in $\{-1,0,1\}$. 
As such, we view this as an ordinal outcome, to  which 
we fit an  ordinal version of a rank-1 symmetric  AME  model 
 using the {\tt ame} command:
\begin{knitrout}\footnotesize
\definecolor{shadecolor}{rgb}{0.969, 0.969, 0.969}\color{fgcolor}\begin{kframe}
\begin{alltt}
\hlstd{fit_cw_R1}\hlkwb{<-}\hlkwd{ame}\hlstd{(Y,Xd,Xn,}\hlkwc{R}\hlstd{=}\hlnum{1}\hlstd{,}\hlkwc{model}\hlstd{=}\hlstr{"ord"}\hlstd{,}\hlkwc{symmetric}\hlstd{=}\hlnum{TRUE}\hlstd{,}\hlkwc{burn}\hlstd{=}\hlnum{1000}\hlstd{,}\hlkwc{nscan}\hlstd{=}\hlnum{100000}\hlstd{,}\hlkwc{odens}\hlstd{=}\hlnum{100}\hlstd{)}
\end{alltt}
\end{kframe}
\end{knitrout}
\begin{figure}
\begin{knitrout}\footnotesize
\definecolor{shadecolor}{rgb}{0.969, 0.969, 0.969}\color{fgcolor}
\includegraphics[width=\maxwidth]{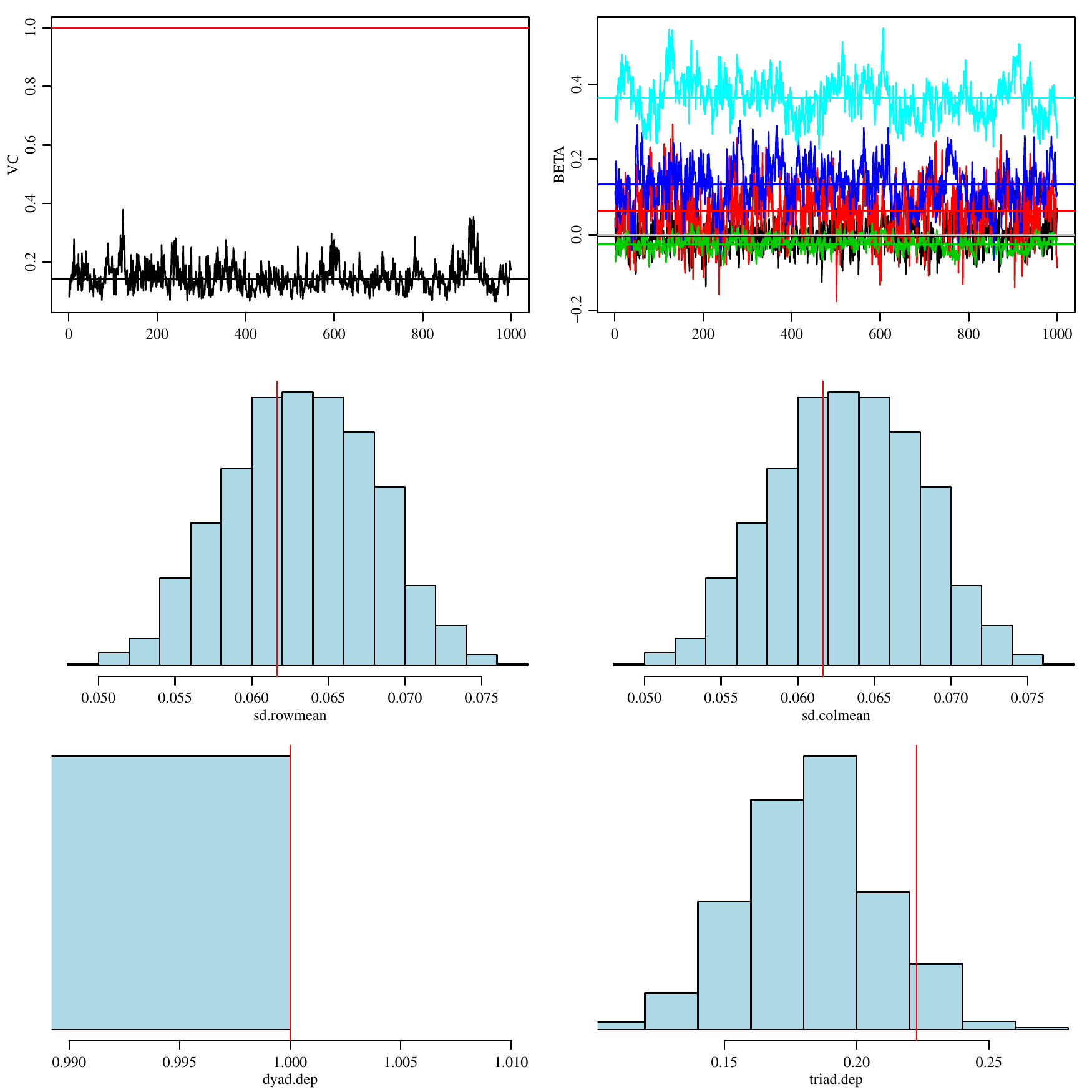} 

\end{knitrout}
\caption{Diagnostic plots for the rank-1 AME model of the {\tt coldwar} data.}
\end{figure}
\noindent
Note that for this symmetric model, 
the row regressors must be the same as the 
column regressors, and so it is sufficient to 
specify these just once. 
We also note that for technical reasons, the mixing  of the MCMC 
algorithm for estimating the low-rank matrix 
$\bl U \bs \Lambda \bl U^T$ is slower than that 
for the asymmetric matrix $\bl U\bl V^T$. For this reason
we lengthened the burn-in period for the Markov chain, and increased 
the number of iterations to 100,000 from the default value of 10,000. 

\begin{knitrout}\footnotesize
\definecolor{shadecolor}{rgb}{0.969, 0.969, 0.969}\color{fgcolor}\begin{kframe}
\begin{alltt}
\hlkwd{summary}\hlstd{(fit_cw_R1)}
\end{alltt}
\begin{verbatim}

Regression coefficients:
             pmean   psd z-stat p-val
lgdp.node   -0.002 0.041 -0.058 0.954
polity.node  0.062 0.070  0.888 0.375
igdp.dyad   -0.025 0.020 -1.276 0.202
ipol.dyad    0.133 0.060  2.211 0.027
ldist.dyad   0.365 0.054  6.745 0.000

Variance parameters:
   pmean   psd
va 0.149 0.046
ve 1.000 0.000
\end{verbatim}
\begin{alltt}
\hlstd{fit_cw_R1}\hlopt{$}\hlstd{L}       \hlcom{# eigenvalue}
\end{alltt}
\begin{verbatim}
[1] 63.61187
\end{verbatim}
\end{kframe}
\end{knitrout}
The results indicate no strong association between 
country-specific levels  of $z_{i,j}$ 
with the nodal attributes. 
At the dyadic level however, 
there appears to be homophily in terms of 
polity. 
Furthermore, 
the parameter for {\tt ldist.dyad} suggests that 
large geographic distance is positively associated 
with cooperation. However,  a better interpretation might be 
that large distance is negatively associated 
with conflict, as most conflicts are regional.
More refined hypotheses abut conflict and 
cooperation could be evaluated by 
fitting separate models 
for the conflict network $(y_{i,j}<0)$  and the 
cooperation network $(y_{i,j}>0)$.

\begin{figure}
\begin{knitrout}\footnotesize
\definecolor{shadecolor}{rgb}{0.969, 0.969, 0.969}\color{fgcolor}
\includegraphics[width=\maxwidth]{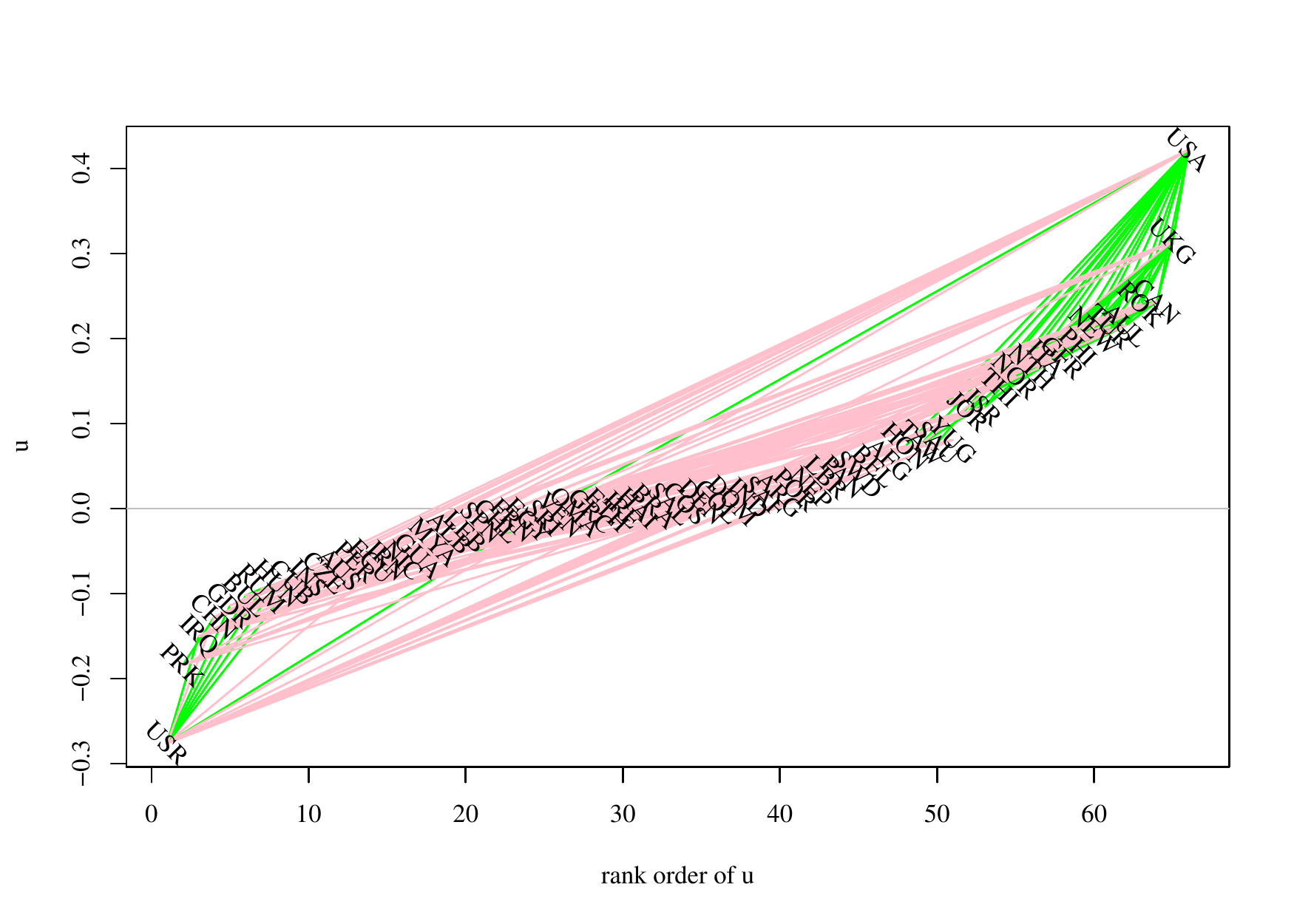} 

\end{knitrout}
\caption{One-dimensional latent factor plot for the {\tt coldwar} analysis.
Green  and red lines indicate cooperation and conflict,
respectively.}
\label{fig:cwfplot}
\end{figure}

We now describe the estimate of
the low-rank latent factor term $\bl U \bs\Lambda \bl U^T$. 
This term describes heterogeneity in the dataset that is not 
explained by the nodal or dyadic regressors, or the terms in the 
social relations covariance model.  
As shown above, 
the estimated ``eigenvalue'' {\tt fit\_cw\$L} is positive.
Since $y_{i,j}$ is increasing in $\bl u_i^T \bs \Lambda \bl u_j$
 (which is  $\lambda u_i u_j$  in this rank-1 model) 
this means that countries that cooperate should 
on average have 
estimated  $\bl u$-vectors pointing in the same 
direction, and countries in conflict 
should have estimates pointing in opposite directions. 
A plot of the latent factors in 
Figure \ref{fig:cwfplot} confirms this, showing that 
cooperative pairs (linked by green lines) 
 essentially all have $u$-values that are on the same side of the origin
(the one exception involves  Egypt, which was cooperative with both the 
USA and USSR). 
Conflictual pairs (linked by red lines)
are generally on opposite sides of the origin.


\bibliography{amen}

\end{document}